\documentclass[%
 reprint,
 amsmath,amssymb,
 aps,
prl,
]{revtex4-2}

\usepackage{graphicx}
\usepackage{dcolumn}
\usepackage{bm}
\usepackage{physics,comment}
\usepackage{xparse}
\usepackage{cleveref}
\usepackage{colortbl}
\usepackage{makecell}

\begin{document}

\preprint{APS/123-QED}

\title{Colloidal Magnus effect in polymer solutions}

\author{Marco De Corato}
\email{mdecorato@unizar.es}
\affiliation{ Aragon Institute of Engineering Research (I3A), University of Zaragoza, Zaragoza, Spain}
\author{Kun Zhang}%
\author{Lailai Zhu}%
\email{lailai\_zhu@nus.edu.sg}
\affiliation{%
Department of Mechanical Engineering, National University of Singapore, Singapore, Singapore}%

\def \omeganon {\Omega^*}
\def \viz {\textit{viz.},~}
\def \ie {\textit{i.e.},~}
\def \eg {\textit{e.g.},~}
\def \mV {\mathcal{V}}
\def \mS {\mathcal{S}}

\date{\today}

\begin{abstract}
Rotating particles moving in fluids undergo a transverse migration
via the inertia-induced Magnus effect. This phenomenon vanishes at colloidal scales because inertia is negligible and the fluid flow is time reversible. 
Yet, recent experiments discovered an inverse Magnus effect of colloids in polymeric and micellar solutions supposedly because their viscoelasticity breaks the time reversibility. 
Our study shows that classical viscoelastic features---normal-stress differences and/or shear-thinning cannot explain this phenomenon. 
Instead, it originates from local polymer density inhomogeneities due to their stress-gradient-induced transport, 
a mechanism increasingly important at smaller scales---indeed relevant to colloidal experiments. 
Incorporating this mechanism 
into 
our model leads to quantitative agreement with the experiments without fitting parameters.
Our work provides new insights into colloidal motion in complex fluids with microstructural inhomogeneities, offers a simple mechanistic theory for predicting the resulting migration, and underscores the necessity of assimilating these findings in future designs of micro-machinery including swimmers, actuators, rheometers, and so on.
\end{abstract}

\maketitle

The viscoelasticity of polymeric fluids enables them to exhibit effects that are often distinct or even contrary to those triggered by weak or intermediate inertia in Newtonian fluids. Consider a macroscopic object forced to move in water; removing the forcing suddenly, it continues to cruise forward due to inertia. Conversely, a colloid in viscoelastic fluids moves backward upon force removal---a phenomenon termed recoil in microrheology~\citep{chapman2014nonlinear}. Another example is the rod-climbing or ``Weissenberg'' effect~\cite{weissenberg1947}, where the free surface of a complex fluid climbs up a spinning rod,
whereas the near-rod surface would bend down in Newtonian fluids with inertia. Similarly, while inertia drives the wind from the poles to the equator at the scale of the rotating Earth, desktop experiments~\citep{giesekus1965some,hill1972nearly} showcased a reversely-directed secondary flow around a rotating sphere in polymeric liquids.

\begin{figure}
	\centering
	\includegraphics[width=1\linewidth]{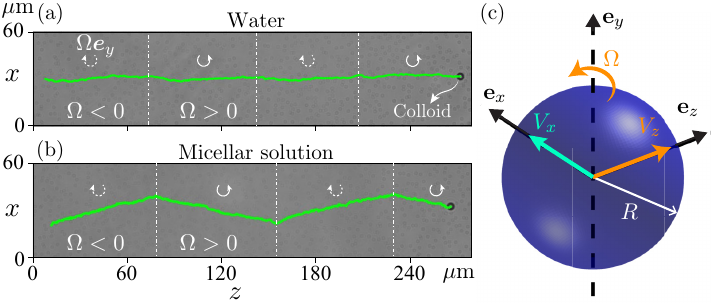}
\caption{Experiments demonstrate that a colloid spinning about $\boldsymbol{e}_y$ and translating along $\boldsymbol{e}_z$ in (a) water exhibits negligible lateral ($\boldsymbol{e}_x$) motion; Contrastingly, it experiences an inverse Magnus effect in (b) a micellar solution, adapted from \citet{cao2023}.
(c), Schematic of our model: a spherical particle with a prescribed translational velocity $V_z$ and rotational velocity $V_x$, with its migration velocity $V_x$ to be determined.
 } 
	\label{fig:sketch}
\end{figure}

Recent experiments~\cite{cao2023} presented another compelling illustration of this contrasting trend. As depicted in Figure \ref{fig:sketch}(b), a spinning and translating micron-sized colloidal particle in micellar solutions with negligible inertia was observed migrating perpendicular to both its spinning and translational axes. Intriguingly, the transversal migration was oriented opposite to that observed in inertial Newtonian fluids---a phenomenon commonly known as the Magnus effect~\citep{newton1993new,magnus1853ueber}.
Such an inverse Magnus migration disappears when driving the colloid in a Newtonian fluid [see Figure \ref{fig:sketch}(a)], as forbidden by the time-reversibility of the Stokes flow. 

Fluid viscoelasticity is known to break the time-reversibility, allowing micron-sized particles to migrate across streamlines within complex fluids.
Since its first observation by \citet{karnis1966particle},
viscoelasticity-induced particulate migration has been widely reported~\cite{ho1976migration,chan1977note,
leal1979motion,brunn1980motion,leshansky2007tunable,
d2017particle,yuan2018recent}. 
Typically, it originates from normal stress differences associated with non-uniform shear rates and/or geometric asymmetries, as exemplified by the lateral motion of particles towards the centerline of a Poiseuille flow and the outer wall of a rotating Couette flow~\cite{ho1976migration,chan1977note,brunn1976slow}. These findings have spurred research into viscoelastic
microfluidics for particle manipulation~\cite{michele1977alignment,leshansky2007tunable, d2012migration, kim2012lateral, van2014string, lim2014lateral, lim2014inertio, seo2014particle, li2015dynamics,lu2017particle, d2017particle, jaensson2018shear,yu2019equilibrium,zhou2020viscoelastic} 
and material property measurements \cite{del2017relaxation}.  
Another example is the horizontal drift of a sedimenting colloid along the flow or gradient direction of a shear flow~\cite{vishnampet2012concentration,einarsson2017spherical,zhang2020lift}, driven by local variations in shear rate induced by the colloid’s translation.
Beyond these scenarios with an ambient flow, normal stress imbalances owing to geometric asymmetry can enable spinning~\cite{pak2012micropropulsion,su2022viscoelastic, kroo2022freely} or swinging~\cite{gagnon2014fluid} objects to propel in quiescent environments. Does the colloidal Magnus migration uncovered by \citet{cao2023} fit into the established categories or share any mechanistic similarities with these phenomena?

In this work, we elucidate the mechanism behind the novel Magnus effect~\cite{cao2023} 
by theoretically and numerically solving the fluid flow and rheological constitutive equations. We find that the normal stress difference and/or shear thinning, typical of polymer solutions, cannot explain the Magnus migration. This finding contrasts with the above-mentioned studies that typically link similar low-Reynolds-number colloidal migration in viscoelastic media to their normal stress differences. Crucially, we identify, through asymptotic analysis and simulations, 
the flow-induced inhomogeneous distribution of polymers as an indispensable enabler for the new Magnus effect---a mechanism 
previously overlooked in colloidal dynamics.
Importantly, our asymptotic prediction of the Magnus effect quantitatively align with experimental data~\cite{cao2023}.

We consider a spherical particle of radius $R$ suspended in a polymer solution. 
The particle is centered within a Cartesian coordinate system, as illustrated in Figure \ref{fig:sketch}(c). It rotates steadily around the $y$-axis with angular velocity $\Omega\boldsymbol{e}_y$ 
while simultaneously translating along the $z$-axis at a constant velocity $V_z$. Due to the Magnus effect, the particle might migrate transversely, \ie in the $x$-direction. This corresponding transversal velocity $V_x$ remains to be determined as part of the solution.

\begin{figure*}[htp!]
	\includegraphics[width=1.0\linewidth]{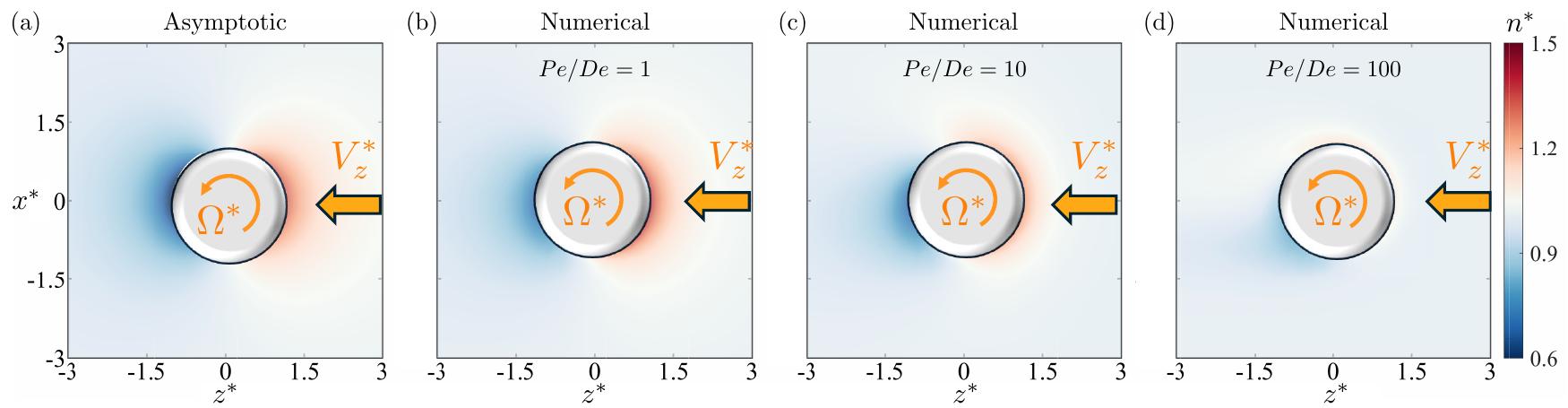}
\caption{Distributions of polymer number density $n^*$ in the symmetry plane $y^*=0$. (a), Asymptotic formula Eq. \eqref{eq:n1_l} at $De=0.3$. (b-c), Numerical results at increasing $Pe/De$ values. When $Pe/De \gg 1$, advection dominates, leading the polymer distribution to approach a homogeneous state. 
The remaining parameters are $\Omega^{*}=0.5$, $De=0.3$, and $\beta=0.1$. 
} 
	\label{fig:np}
\end{figure*} 
By neglecting the particle and fluid inertia, the governing equations for the velocity $\boldsymbol{v}$ and pressure $p$ of the polymeric flow are
\begin{equation}\label{mombal}
\grad\cdot \boldsymbol{v} = 0, \quad 
    \grad \cdot \left( 2 \eta_s \boldsymbol{D} -p\boldsymbol{I} + \boldsymbol{\tau}  \right)= \boldsymbol{0} \,\, , 
\end{equation}
where $\eta_s$ is the solvent viscosity, 
$\boldsymbol{D}=\left[\grad \boldsymbol{v}+\left( \grad  \boldsymbol{v}\right)^T \right]/2$ the strain-rate tensor, and $\boldsymbol{\tau}$ the polymeric stress.   
Far from the particle, the fluid velocity  $\boldsymbol{v}=-V_z \boldsymbol{e}_z-V_x \boldsymbol{e}_x$.
At the surface $\mS$ of the particle, $\boldsymbol{v}= \Omega \, \boldsymbol{e}_y \times \boldsymbol{r}$, where $\boldsymbol{r}$ is the position vector.
To determine the migration velocity $V_x$,  we apply the force-free condition that the total hydrodynamic force exerted on the particle vanishes in the $x$-direction, namely, $\int_{\mS} \left( 2 \eta_s \boldsymbol{D} -p\boldsymbol{I} + \boldsymbol{\tau}  \right) : \boldsymbol{n}\boldsymbol{e}_x d\mS=0$, where 
$\boldsymbol{n}$ denotes the fluid-pointing unit normal.

To model the polymeric flow, we first attempt the widely-employed Oldroyd-B (OB) constitutive equation \cite{bird1987dynamics,larson2013constitutive,Morozov2015}, which represents polymers as linear elastic dumbbells with a spring constant $k$. The polymers are homogeneously distributed in a Newtonian solvent, with a uniform number density $n = n_{\infty}$, where $n_{\infty}$ is the constant far-field density. 
The polymeric stress $\boldsymbol{\tau}= k \, \boldsymbol{C}-k_BT n \boldsymbol{I}$ scales linearly with the deviation of the density-weighted chain conformation $\boldsymbol{C}$ from its thermal equilibrium. Here, $k_B$ indicates the Boltzmann constant, and $T$ is the absolute temperature. The governing equation for $\boldsymbol{C}$ is~\cite{apostolakis2002stress}:
\begin{equation}\label{conftranspeq}
       \lambda \, \overset{\nabla}{\boldsymbol{C}} + \boldsymbol{C}  - \frac{k_BT}{k} \, n \, \boldsymbol{I} =\boldsymbol{0} \, \, ,
\end{equation}
with $\lambda$ the relaxation time of the polymers and the upper convected derivative defined as $\overset{\nabla}{\boldsymbol{C}} = \frac{\partial \boldsymbol{C}}{\partial t} +  \boldsymbol{v} \cdot \grad \boldsymbol{C} - \left(\grad\boldsymbol{v}\right)^T \cdot \boldsymbol{C} -\boldsymbol{C} \cdot \grad \boldsymbol{v}$.

Scaling lengths, velocities, and time by $R$, $V_z$, and $V_z/R$, respectively, we non-dimensionalize the governing  equations, yielding three dimensionless parameters:  (i) the Deborah number $De=\lambda V_z/R$ characterizing the polymer relaxation time $\lambda$ to the rate of deformation; (ii) the ratio of the solvent viscosity $\eta_s$ to the total viscosity, $\beta = \eta_s/(\eta_s+\eta_{p,\infty})$,  where $\eta_{p,\infty}=n_\infty \lambda k_BT$ is the far-field polymeric viscosity; 
and (iii) the ratio of the rotational tangential velocity (from colloidal rotation) to the translational velocity, $\omeganon=\Omega R/V_z$. 
Dimensionless variables are marked with asterisks hereafter.

Since the experiments showed a quadratic relationship $V_x \propto -\Omega V_z$ \cite{cao2023}, we expect a perturbation expansion of Eqs. \labelcref{mombal,conftranspeq}, valid for $De \ll 1$, to yield a similar scaling. To zeroth-order in $De$, the flow and stress fields are linearly proportional to the velocity, and the Magnus effect is forbidden. To first order in $De$, these fields are quadratic in the velocity and we may expect a nonzero migration velocity $V_x$. By performing the expansion up to this order, we obtain $V_x^* = De \, V_{x,1}^*+ \mathcal{O}(De^2)$, and then determine the first-order correction $ V_{x,1}^*$ using the Lorentz reciprocal theorem \cite{leal1979motion,masoud2019reciprocal},
see End Matter (EM). Surprisingly, we find $V_{x,1}^*=0$, consistent with numerical solutions to Eqs.~\labelcref{mombal,conftranspeq} obtained by a finite element method [FEM, see (EM)],  indicating no transversal migration.
We conclude that the OB model---encompassing viscoelastic memory and the first normal stress difference---cannot account for the Magnus effect observed experimentally. Employing the Giesekus model \cite{larson2013constitutive} that extends the OB model by incorporating a second normal stress difference and shear-thinning properties, also fails to explain the colloidal migration (see EM).

This discrepancy with the experiments prompts a reevaluation of whether the assumptions inherent in the OB model align with the experimental conditions.
Specifically, we reconsider the presumed homogeneous distribution of macromolecules.
The original OB model assumes that the transport of macromolecules with a diffusion coefficient of $D$ is dominated by advection, namely, the P\'eclet number $Pe=V_z R/D \rightarrow \infty $. 
In the large $Pe$ limit,
advection homogenizes the macromolecular distribution. 
However, this might not be true for flows past micron-sized objects such as those used in the experiments. At small scales, the polymers can be sufficiently extended to feel the curvature of the streamlines around the colloid and the resulting gradients of stress drive the transport of polymers~\cite{aubert1980macromolecules,helfand1989large,doi1990effects,milner1991hydrodynamics,mavrantzas1992modeling}. 
If the rate of stress-gradient-induced transport is faster or comparable to advection, then
the polymers will be distributed inhomogeneously.  The relative importance of advection to the stress-gradient-induced transport is quantified by the ratio $Pe/De = R^2/\lambda D$ \cite{tsouka2014stress}. 
For a given polymer solution, this ratio depends on the particle size only. 
For $Pe/De \gg 1$, advection dominates and the OB model is applicable, while for $Pe/De \lessapprox 1$, we expect considerable polymer inhomogeneities. By using a characteristic value of diffusion coefficient $D \approx 10^{-12} \, \rm{m^2 \, s^{-1}}$ 
and the experimental values $R \approx 2 \, \rm{\mu m}$ and $\lambda \approx 1-10\, \rm{s}$ \cite{gomez2015transient, cao2023}, we estimate $Pe/De \approx 4-0.4$, indeed suggesting that stress-gradient-induced transport of polymers is comparable or even stronger than advection in the experiments. 

Consequently, we incorporate this effect following
Refs.~\cite{apostolakis2002stress,tsouka2014stress}. 
We treat the number density of polymers, $n$, as a variable that satisfies 
\begin{equation}\label{dumbtransp}
\frac{\partial \, n}{\partial \, t} + \boldsymbol{v} \cdot \grad n = D \grad \cdot \left( \grad n - \frac{1}{k_{B}T}\grad \cdot \boldsymbol{\tau} \right) \, \, ,
\end{equation}
where the rightmost term models the flux of polymer molecules driven by the gradient of the polymeric stress \cite{apostolakis2002stress}. Accordingly, the polymer viscosity $\eta_p = n \lambda k_B T$ varies spatio-temporally. The far-field density remains as
$n=n_\infty$, while polymers cannot penetrate the particle surface, where 
$  \left( \grad n - \grad \cdot \boldsymbol{\tau}/k_BT \right) \cdot \boldsymbol{n}=0.$
We make the number density dimensionless using $n_\infty$, \ie $n^{*}=n/n_{\infty}$ and report the full dimensionless equations in the EM.

\begin{figure*}[htp!]
	\includegraphics[width=1\linewidth]
{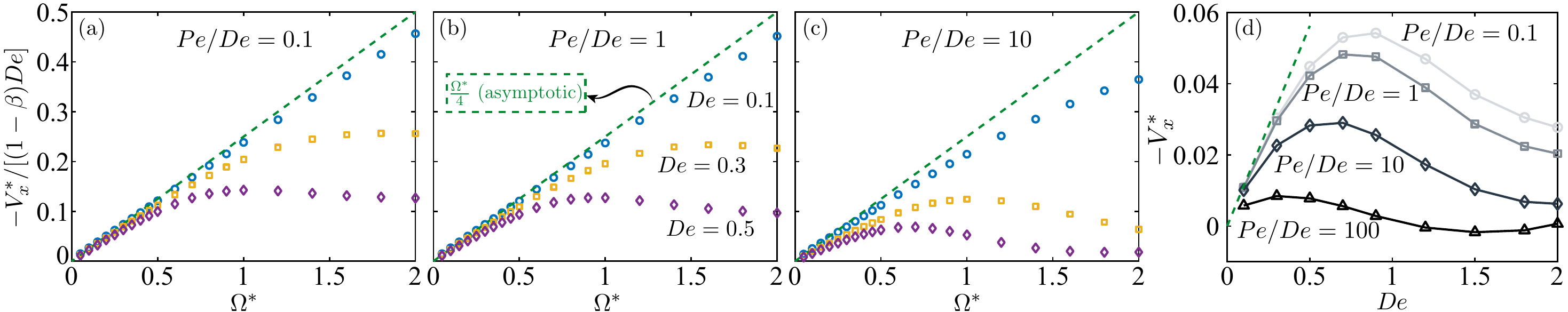}
\caption{
Numerical Magnus velocity 
$-V^*_x $ 
compared to the asymptotic prediction Eq.~\eqref{Magnus_effect_dimless} versus the particle's dimensionless spinning rate $\omeganon$, when the ratio of advection to stress-gradient-induced transport
is (a) weak, $Pe/De=0.1$; (b) intermediate, $Pe/De=1$; and (c) strong, $Pe/De=10$.  (d), $-V_x^*$ versus $De$ at four $Pe/De$ values when $\omeganon = 0.5$. Here, $\beta=0.1$.
 } 
 \label{fig:rotvel}
\end{figure*}

We perform a perturbation expansion of 
Eqs.~\labelcref{mombal,conftranspeq,dumbtransp}, valid for small $De$ and $Pe$ numbers (see EM). 
To first order in $De$ and $Pe$, we identify an inhomogeneous distribution of number density,
\begin{equation}\label{eq:n1_l}
n^* = 1+\frac{3 De  \, z^*}{2 {(r^*)}^3} +\mathcal{O}(De^2, De \, Pe) \, \, ,
\end{equation}
where $r^*$ is the dimensionless distance from the origin. Notably, the first-order density deviation, $3De z^*/2 {(r^*)}^3$, from the far-field state $n^*_{\infty}=1$ depends solely on $De$, and is independent of the rotational rate, $\Omega$. 
Linearly proportional to $z^*$, this deviation indicates
a dipolar distribution of polymers, revealing their accumulation
at the front ($z^*>0$) 
and depletion at the rear ($z^*<0$) of the moving particle, see Figure~\ref{fig:np}(a). 
This distribution is confirmed by FEM-based numerical solutions (see EM) to Eqs.~\labelcref{mombal,conftranspeq,dumbtransp}, as illustrated in Figure~\ref{fig:np}. 
The asymptotic solution, Eq. \eqref{eq:n1_l} agrees with simulations up to $Pe/De \approx 10$ for $De=0.3$ and $\Omega^*=0.5$. Remarkably, the perturbation expansion remains valid even when 
flow advection exceeds stress-gradient-induced transport of polymers. At large $Pe/De$, the concentration dipole tilts away from the colloid's translational orientation ($\boldsymbol{e}_z$) [see Figure~\ref{fig:np}(c-d)],
a signature of growing rotational convection effect---as also identified by \citet{cao2023}.

By further extending the asymptotic analysis through the Lorentz reciprocal theorem, we derive the leading-order transversal velocity of the colloid, signifying the Magnus effect (see EM):
\begin{equation}\label{Magnus_effect_dimless}
    V_x^*=-\frac{(1-\beta) \, De \, \Omega^*}{4}  
\end{equation}
with its dimensional form:
\begin{equation}\label{Magnus_effect_dim}
    V_x=-\frac{\lambda \, \eta_{p,\infty} \, V_z \, \Omega}{4(\eta_s +\eta_{p,\infty}) }.
\end{equation}
This asymptotic prediction captures the experimentally identified trend, $V_x/V_z = -\alpha \, \Omega$ (where $\alpha>0$)~\cite{cao2023}, in both sign and scaling, as further examined below. Moreover, Eq.~\eqref{Magnus_effect_dimless} is cross-validated against FEM simulations, as evidenced in Figure~\ref{fig:rotvel}. In the weak rotation limit, $\Omega^* \ll 1$, the numerical and theoretical data match closely for small $De$ (\eg $0.1$) and $Pe/De$ (\eg $0.1$ and $1$). However, deviations emerge when $De \geq 0.3$, and systematically grow with $\Omega^*$. Evidently, as the colloid spins faster, the numerical results fall increasingly below the linear asymptotic trend. Specifically at $De=0.3$, they approach a saturated value when $Pe/De=0.1$ [Figure~\ref{fig:rotvel}(a)] or $1$ [Figure~\ref{fig:rotvel}(b)], resonating the velocity saturation reported in the experiments~\cite{cao2023}. 

Additionally, Figure~\ref{fig:rotvel}(d) reveals the Magnus velocity $-V_x^*$ versus $De$ at a spinning rate of $\omeganon = 0.5$. The velocity decreases with increasing $Pe/De$, eventually becoming negligible when $Pe/De=100$. Indeed, in this regime of dominant advection, the OB model is recovered, resulting in no colloid migration, as we find in our initial attempt.
In the opposite limit $Pe/De=0.1$ (weak advection), simulations quantitatively reproduce the asymptotic linear dependence of $-V_x^*$ on $De$ at low values of $De$. Beyond that regime, the velocity changes non-monotonically with $De$ and peaks at $De\approx 1$. Notably, such a peak was reported in studies of micro-locomotion in viscoelastic fluids~\cite{teran2010viscoelastic,liu2011force}.

\begin{figure}[htp!]
	\centering
	\includegraphics[width=1.0\linewidth]{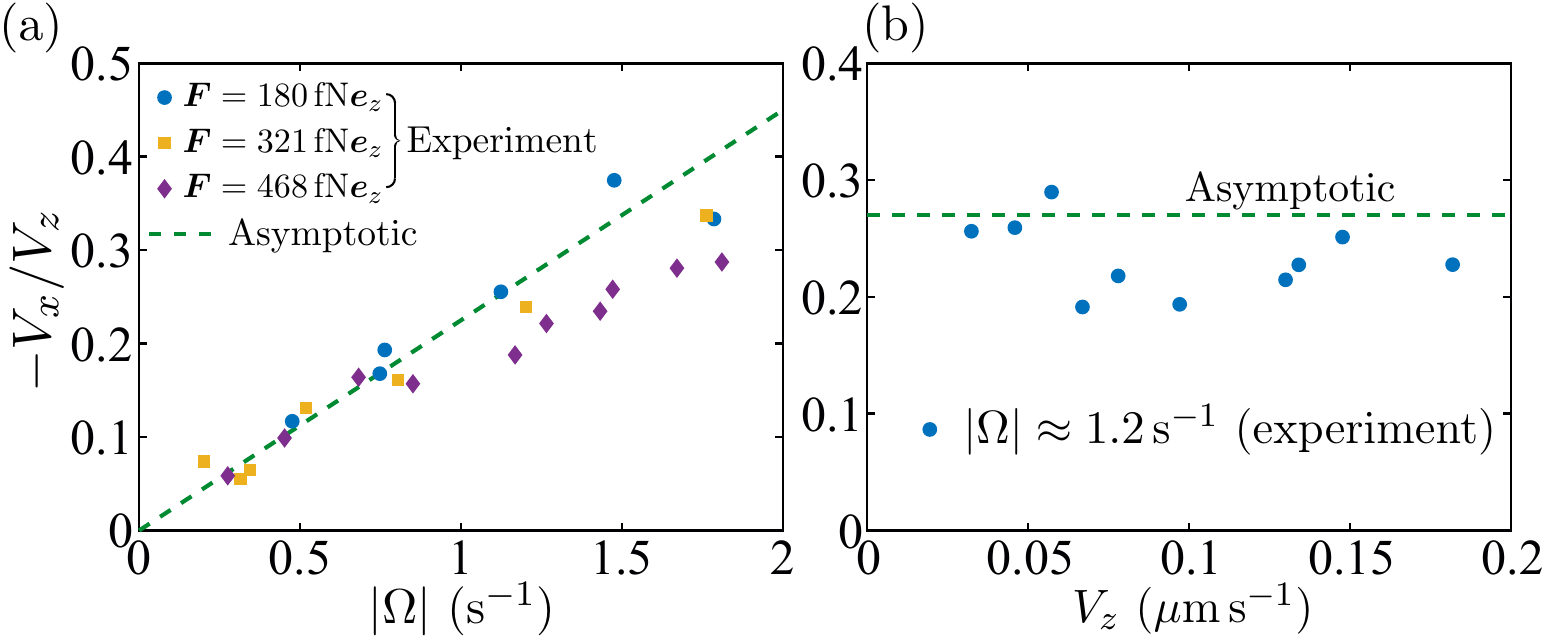}
\caption{Quantitative agreement (using no fitting parameters) between our asymptotic prediction, Eq. \eqref{Magnus_effect_dim}, and the experimental results~\cite{cao2023} for a rotating trimer in a micellar solution, see Figure~\ref{fig:sketch}(b). It experienced  (a) three $z$-oriented forces $\boldsymbol{F}$ with various rotating speeds $|\Omega|$, and (b) varying translational velocities $V_z$ at a prescribed $|\Omega|\approx 1.2\,\rm{s}^{-1}$.   
}
	\label{fig:comparison}
\end{figure}

We now apply the numerically-validated theory to characterize the experimental observations of 
\citet{cao2023} on colloidal trimers in a micellar solution. Its rheological properties, measured and previously reported by the same group~\cite{gomez2015transient}, make
this characterization possible. Particularly, these properties include $\lambda \approx 0.9\,\rm{s}$ (see EM)
and $\eta_{p,\infty}/\left( \eta_s + \eta_{p,\infty} \right) \rightarrow 1$ (see EM). 
Substituting them into Eq.~\eqref{Magnus_effect_dim}, a quantitative agreement emerges between the theory and experiments in Figure~\ref{fig:comparison}(a), where the trimers were forced to translate at three speeds and spin at various rates. A similar agreement holds for a trimer spinning at $|\Omega|\approx 1.2 \,\text{s}^{-1}$ but translating at various speeds, as depicted in Figure~\ref{fig:comparison}(b). Remarkably, this agreement does not rely on any fitting parameters. 
Admittedly, the asymptotic prediction overestimates the Magnus effect of strong spinners (large $\Omega$)---a discrepancy that grows more pronounced at higher translational velocities $V_z$ (subject to the increased $\boldsymbol{e}_z$-oriented force $\boldsymbol{F}$). In fact, numerical data in Figure~\ref{fig:rotvel}(a-c) have revealed a similar deviation, where a larger $De=\lambda V_z/R$ in simulations corresponds to a higher $V_z$ in experiments. Indeed, as $De$ increases, higher-order effects become increasingly important, leading to larger deviations.
Note that in the experiment the authors used a trimer, while we consider a sphere here.

Based on our findings, we attribute 
the experimentally observed Magnus effect to two mechanisms. 
First, the colloidal translation and the concomitant stress-gradient-induced transport 
cause the polymer concentration $n$ to be higher at the front but lower at the rear of the colloid, yielding a dipolar distribution of polymeric viscosity $\eta_p = n \lambda k_B T$ around it.
Second, the colloidal rotation in such a viscosity profile results in a fore-aft stress imbalance, driving the transversal 
migration. Indeed, the migration velocity, Eq. \eqref{Magnus_effect_dimless}, is identical to that of a sphere rotating in a Newtonian fluid with a dipolar distribution of viscosity \cite{oppenheimer2016motion}. 
This analogy indirectly suggests that the Magnus effect stems purely from the spatial viscosity variation due to an inhomogeneous polymer distribution. 
Normal stress differences and shear-thinning might play a secondary role, likely becoming important at much higher $De$ numbers. These findings align with recent studies highlighting the importance of
local microstructure and viscosity inhomogeneities 
in the dynamics of active particles \cite{liebchen2018viscotaxis,datt2019active, dandekar2020swimming, shaik2021hydrodynamics,esparza2021dynamics,stehnach2021viscophobic, olsen2021active,de2025enhanced}. 

\textit{Conclusions}---We have established that the inverse Magnus migration of a colloid, experimentally observed in polymeric solutions~\cite{cao2023}, 
cannot be explained by normal stress differences or shear-thinning rheology. 
Instead, our analysis reveals that stress-gradient-induced polymer transport and the ensuing local viscosity variations underlie the observed migration. This mechanistic interpretation is validated by the striking quantitative agreement between experimental data and our theoretical model, which incorporate these mechanisms---notably, without adjustable parameters. 
Critically, our results underscore the necessity of reassessing standard constitutive models predicated on homogeneous polymer distributions, both for modeling colloidal dynamics in complex fluids and, more broadly, in (sub)micro-scale polymeric flows.
In these cases, macromolecules can feel the curvature of the streamlines leading to significant 
concentration inhomogeneities, which can play a more important role than viscoelastic stresses.

\begin{acknowledgments}
MDC was supported by the Ramon y Cajal fellowship RYC2021-030948-I and by the PID2022-139803NB-I00  research grant funded by the MICIU/AEI /10.13039/501100011033 and by the EU under the NextGenerationEU/PRTR program.
LZ acknowledges the support from the Singapore Ministry of Education Academic Research Fund Tier 2 grant (MOE-T2EP50221-0012 and MOE-T2EP50122-0015).
Some computation of the work was performed on resources of the National Supercomputing Centre, Singapore (https://www.nscc.sg).
We thank Xin Cao for initial discussions and
Hua Zhang for performing some preliminary simulations. 
MDC and LZ thank Eric Keaveny, Blaise Delmotte, and the European Mechanics Society for the invitation to the 645 EUROMECH Colloquium on Nonlinear Dynamics at Zero Reynolds Numbers, which fostered this collaboration.
\end{acknowledgments}


\begin{thebibliography}{59}%
\makeatletter
\providecommand \@ifxundefined [1]{%
 \@ifx{#1\undefined}
}%
\providecommand \@ifnum [1]{%
 \ifnum #1\expandafter \@firstoftwo
 \else \expandafter \@secondoftwo
 \fi
}%
\providecommand \@ifx [1]{%
 \ifx #1\expandafter \@firstoftwo
 \else \expandafter \@secondoftwo
 \fi
}%
\providecommand \natexlab [1]{#1}%
\providecommand \enquote  [1]{``#1''}%
\providecommand \bibnamefont  [1]{#1}%
\providecommand \bibfnamefont [1]{#1}%
\providecommand \citenamefont [1]{#1}%
\providecommand \href@noop [0]{\@secondoftwo}%
\providecommand \href [0]{\begingroup \@sanitize@url \@href}%
\providecommand \@href[1]{\@@startlink{#1}\@@href}%
\providecommand \@@href[1]{\endgroup#1\@@endlink}%
\providecommand \@sanitize@url [0]{\catcode `\\12\catcode `\$12\catcode
  `\&12\catcode `\#12\catcode `\^12\catcode `\_12\catcode `\%12\relax}%
\providecommand \@@startlink[1]{}%
\providecommand \@@endlink[0]{}%
\providecommand \url  [0]{\begingroup\@sanitize@url \@url }%
\providecommand \@url [1]{\endgroup\@href {#1}{\urlprefix }}%
\providecommand \urlprefix  [0]{URL }%
\providecommand \Eprint [0]{\href }%
\providecommand \doibase [0]{https://doi.org/}%
\providecommand \selectlanguage [0]{\@gobble}%
\providecommand \bibinfo  [0]{\@secondoftwo}%
\providecommand \bibfield  [0]{\@secondoftwo}%
\providecommand \translation [1]{[#1]}%
\providecommand \BibitemOpen [0]{}%
\providecommand \bibitemStop [0]{}%
\providecommand \bibitemNoStop [0]{.\EOS\space}%
\providecommand \EOS [0]{\spacefactor3000\relax}%
\providecommand \BibitemShut  [1]{\csname bibitem#1\endcsname}%
\let\auto@bib@innerbib\@empty
\bibitem [{\citenamefont {Chapman}\ and\ \citenamefont
  {Robertson-Anderson}(2014)}]{chapman2014nonlinear}%
  \BibitemOpen
  \bibfield  {author} {\bibinfo {author} {\bibfnamefont {C.~D.}\ \bibnamefont
  {Chapman}}\ and\ \bibinfo {author} {\bibfnamefont {R.~M.}\ \bibnamefont
  {Robertson-Anderson}},\ }\bibfield  {title} {\bibinfo {title} {Nonlinear
  microrheology reveals entanglement-driven molecular-level viscoelasticity of
  concentrated {DNA}},\ }\href@noop {} {\bibfield  {journal} {\bibinfo
  {journal} {Phys. Rev. Lett.}\ }\textbf {\bibinfo {volume} {113}},\ \bibinfo
  {pages} {098303} (\bibinfo {year} {2014})}\BibitemShut {NoStop}%
\bibitem [{\citenamefont {Weissenberg}(1947)}]{weissenberg1947}%
  \BibitemOpen
  \bibfield  {author} {\bibinfo {author} {\bibfnamefont {K.}~\bibnamefont
  {Weissenberg}},\ }\bibfield  {title} {\bibinfo {title} {A continuum theory of
  rheological phenomena},\ }\href@noop {} {\bibfield  {journal} {\bibinfo
  {journal} {Nature}\ }\textbf {\bibinfo {volume} {159}},\ \bibinfo {pages}
  {310} (\bibinfo {year} {1947})}\BibitemShut {NoStop}%
\bibitem [{\citenamefont {Giesekus}(1965)}]{giesekus1965some}%
  \BibitemOpen
  \bibfield  {author} {\bibinfo {author} {\bibfnamefont {H.}~\bibnamefont
  {Giesekus}},\ }\bibfield  {title} {\bibinfo {title} {Some secondary flow
  phenomena in general viscoelastic fluids},\ }in\ \href@noop {} {\emph
  {\bibinfo {booktitle} {Proc. 4th Intern. Congr. Rheology, Part 1}}}\
  (\bibinfo {year} {1965})\BibitemShut {NoStop}%
\bibitem [{\citenamefont {Hill}(1972)}]{hill1972nearly}%
  \BibitemOpen
  \bibfield  {author} {\bibinfo {author} {\bibfnamefont {C.~T.}\ \bibnamefont
  {Hill}},\ }\bibfield  {title} {\bibinfo {title} {Nearly viscometric flow of
  viscoelastic fluids in the disk and cylinder system. {II}: Experimental},\
  }\href@noop {} {\bibfield  {journal} {\bibinfo  {journal} {Trans. Soc.
  Rheol.}\ }\textbf {\bibinfo {volume} {16}},\ \bibinfo {pages} {213} (\bibinfo
  {year} {1972})}\BibitemShut {NoStop}%
\bibitem [{\citenamefont {Cao}\ \emph {et~al.}(2023)\citenamefont {Cao},
  \citenamefont {Das}, \citenamefont {Windbacher}, \citenamefont {Ginot},
  \citenamefont {Kr\"uger},\ and\ \citenamefont {Bechinger}}]{cao2023}%
  \BibitemOpen
  \bibfield  {author} {\bibinfo {author} {\bibfnamefont {X.}~\bibnamefont
  {Cao}}, \bibinfo {author} {\bibfnamefont {D.}~\bibnamefont {Das}}, \bibinfo
  {author} {\bibfnamefont {N.}~\bibnamefont {Windbacher}}, \bibinfo {author}
  {\bibfnamefont {F.}~\bibnamefont {Ginot}}, \bibinfo {author} {\bibfnamefont
  {M.}~\bibnamefont {Kr\"uger}},\ and\ \bibinfo {author} {\bibfnamefont
  {C.}~\bibnamefont {Bechinger}},\ }\bibfield  {title} {\bibinfo {title}
  {Memory-induced {M}agnus effect},\ }\href@noop {} {\bibfield  {journal}
  {\bibinfo  {journal} {Nat. Phys.}\ }\textbf {\bibinfo {volume} {19}},\
  \bibinfo {pages} {1904} (\bibinfo {year} {2023})}\BibitemShut {NoStop}%
\bibitem [{\citenamefont {Newton}(1993)}]{newton1993new}%
  \BibitemOpen
  \bibfield  {author} {\bibinfo {author} {\bibfnamefont {I.}~\bibnamefont
  {Newton}},\ }\bibfield  {title} {\bibinfo {title} {A new theory about light
  and colors},\ }\href@noop {} {\bibfield  {journal} {\bibinfo  {journal} {Am.
  J. Phys.}\ }\textbf {\bibinfo {volume} {61}},\ \bibinfo {pages} {108}
  (\bibinfo {year} {1993})}\BibitemShut {NoStop}%
\bibitem [{\citenamefont {Magnus}(1853)}]{magnus1853ueber}%
  \BibitemOpen
  \bibfield  {author} {\bibinfo {author} {\bibfnamefont {G.}~\bibnamefont
  {Magnus}},\ }\bibfield  {title} {\bibinfo {title} {Ueber die {Abweichung} der
  {Geschosse}, und: {Ueber} eine auffallende {Erscheinung} bei rotirenden
  k{\"o}rpern},\ }\href@noop {} {\bibfield  {journal} {\bibinfo  {journal}
  {Ann. Phys. (Berl.)}\ }\textbf {\bibinfo {volume} {164}},\ \bibinfo {pages}
  {1} (\bibinfo {year} {1853})}\BibitemShut {NoStop}%
\bibitem [{\citenamefont {Karnis}\ and\ \citenamefont
  {Mason}(1966)}]{karnis1966particle}%
  \BibitemOpen
  \bibfield  {author} {\bibinfo {author} {\bibfnamefont {A.}~\bibnamefont
  {Karnis}}\ and\ \bibinfo {author} {\bibfnamefont {S.~G.}\ \bibnamefont
  {Mason}},\ }\bibfield  {title} {\bibinfo {title} {Particle motions in sheared
  suspensions. {XIX}. {Viscoelastic} media},\ }\href@noop {} {\bibfield
  {journal} {\bibinfo  {journal} {Trans. Soc. Rheol.}\ }\textbf {\bibinfo
  {volume} {10}},\ \bibinfo {pages} {571} (\bibinfo {year} {1966})}\BibitemShut
  {NoStop}%
\bibitem [{\citenamefont {Ho}\ and\ \citenamefont
  {Leal}(1976)}]{ho1976migration}%
  \BibitemOpen
  \bibfield  {author} {\bibinfo {author} {\bibfnamefont {B.~P.}\ \bibnamefont
  {Ho}}\ and\ \bibinfo {author} {\bibfnamefont {G.}~\bibnamefont {Leal}},\
  }\bibfield  {title} {\bibinfo {title} {Migration of rigid spheres in a
  two-dimensional unidirectional shear flow of a second-order fluid},\
  }\href@noop {} {\bibfield  {journal} {\bibinfo  {journal} {J. Fluid Mech.}\
  }\textbf {\bibinfo {volume} {76}},\ \bibinfo {pages} {783} (\bibinfo {year}
  {1976})}\BibitemShut {NoStop}%
\bibitem [{\citenamefont {Chan}\ and\ \citenamefont
  {Leal}(1977)}]{chan1977note}%
  \BibitemOpen
  \bibfield  {author} {\bibinfo {author} {\bibfnamefont {P.~C.-H.}\
  \bibnamefont {Chan}}\ and\ \bibinfo {author} {\bibfnamefont {L.~G.}\
  \bibnamefont {Leal}},\ }\bibfield  {title} {\bibinfo {title} {A note on the
  motion of a spherical particle in a general quadratic flow of a second-order
  fluid},\ }\href@noop {} {\bibfield  {journal} {\bibinfo  {journal} {J. Fluid
  Mech.}\ }\textbf {\bibinfo {volume} {82}},\ \bibinfo {pages} {549} (\bibinfo
  {year} {1977})}\BibitemShut {NoStop}%
\bibitem [{\citenamefont {Leal}(1979)}]{leal1979motion}%
  \BibitemOpen
  \bibfield  {author} {\bibinfo {author} {\bibfnamefont {L.~G.}\ \bibnamefont
  {Leal}},\ }\bibfield  {title} {\bibinfo {title} {The motion of small
  particles in non-{Newtonian} fluids},\ }\href@noop {} {\bibfield  {journal}
  {\bibinfo  {journal} {J Non-Newton. Fluid Mech.}\ }\textbf {\bibinfo {volume}
  {5}},\ \bibinfo {pages} {33} (\bibinfo {year} {1979})}\BibitemShut {NoStop}%
\bibitem [{\citenamefont {Brunn}(1980)}]{brunn1980motion}%
  \BibitemOpen
  \bibfield  {author} {\bibinfo {author} {\bibfnamefont {P.}~\bibnamefont
  {Brunn}},\ }\bibfield  {title} {\bibinfo {title} {The motion of rigid
  particles in viscoelastic fluids},\ }\href@noop {} {\bibfield  {journal}
  {\bibinfo  {journal} {J. Non-Newtonian Fluid Mech.}\ }\textbf {\bibinfo
  {volume} {7}},\ \bibinfo {pages} {271} (\bibinfo {year} {1980})}\BibitemShut
  {NoStop}%
\bibitem [{\citenamefont {Leshansky}\ \emph {et~al.}(2007)\citenamefont
  {Leshansky}, \citenamefont {Bransky}, \citenamefont {Korin},\ and\
  \citenamefont {Dinnar}}]{leshansky2007tunable}%
  \BibitemOpen
  \bibfield  {author} {\bibinfo {author} {\bibfnamefont {A.~M.}\ \bibnamefont
  {Leshansky}}, \bibinfo {author} {\bibfnamefont {A.}~\bibnamefont {Bransky}},
  \bibinfo {author} {\bibfnamefont {N.}~\bibnamefont {Korin}},\ and\ \bibinfo
  {author} {\bibfnamefont {U.}~\bibnamefont {Dinnar}},\ }\bibfield  {title}
  {\bibinfo {title} {Tunable nonlinear viscoelastic ``focusing''' in a
  microfluidic device},\ }\href@noop {} {\bibfield  {journal} {\bibinfo
  {journal} {Phys. Rev. Lett.}\ }\textbf {\bibinfo {volume} {98}},\ \bibinfo
  {pages} {234501} (\bibinfo {year} {2007})}\BibitemShut {NoStop}%
\bibitem [{\citenamefont {D'Avino}\ \emph {et~al.}(2017)\citenamefont
  {D'Avino}, \citenamefont {Greco},\ and\ \citenamefont
  {Maffettone}}]{d2017particle}%
  \BibitemOpen
  \bibfield  {author} {\bibinfo {author} {\bibfnamefont {G.}~\bibnamefont
  {D'Avino}}, \bibinfo {author} {\bibfnamefont {F.}~\bibnamefont {Greco}},\
  and\ \bibinfo {author} {\bibfnamefont {P.~L.}\ \bibnamefont {Maffettone}},\
  }\bibfield  {title} {\bibinfo {title} {Particle migration due to
  viscoelasticity of the suspending liquid and its relevance in microfluidic
  devices},\ }\href@noop {} {\bibfield  {journal} {\bibinfo  {journal} {Annu.
  Rev. Fluid Mech.}\ }\textbf {\bibinfo {volume} {49}},\ \bibinfo {pages} {341}
  (\bibinfo {year} {2017})}\BibitemShut {NoStop}%
\bibitem [{\citenamefont {Yuan}\ \emph {et~al.}(2018)\citenamefont {Yuan},
  \citenamefont {Zhao}, \citenamefont {Yan}, \citenamefont {Tang},
  \citenamefont {Alici}, \citenamefont {Zhang},\ and\ \citenamefont
  {Li}}]{yuan2018recent}%
  \BibitemOpen
  \bibfield  {author} {\bibinfo {author} {\bibfnamefont {D.}~\bibnamefont
  {Yuan}}, \bibinfo {author} {\bibfnamefont {Q.}~\bibnamefont {Zhao}}, \bibinfo
  {author} {\bibfnamefont {S.}~\bibnamefont {Yan}}, \bibinfo {author}
  {\bibfnamefont {S.-Y.}\ \bibnamefont {Tang}}, \bibinfo {author}
  {\bibfnamefont {G.}~\bibnamefont {Alici}}, \bibinfo {author} {\bibfnamefont
  {J.}~\bibnamefont {Zhang}},\ and\ \bibinfo {author} {\bibfnamefont
  {W.}~\bibnamefont {Li}},\ }\bibfield  {title} {\bibinfo {title} {Recent
  progress of particle migration in viscoelastic fluids},\ }\href@noop {}
  {\bibfield  {journal} {\bibinfo  {journal} {Lab Chip}\ }\textbf {\bibinfo
  {volume} {18}},\ \bibinfo {pages} {551} (\bibinfo {year} {2018})}\BibitemShut
  {NoStop}%
\bibitem [{\citenamefont {Brunn}(1976)}]{brunn1976slow}%
  \BibitemOpen
  \bibfield  {author} {\bibinfo {author} {\bibfnamefont {P.}~\bibnamefont
  {Brunn}},\ }\bibfield  {title} {\bibinfo {title} {The slow motion of a sphere
  in a second-order fluid},\ }\href@noop {} {\bibfield  {journal} {\bibinfo
  {journal} {Rheol. Acta}\ }\textbf {\bibinfo {volume} {15}},\ \bibinfo {pages}
  {163} (\bibinfo {year} {1976})}\BibitemShut {NoStop}%
\bibitem [{\citenamefont {Michele}\ \emph {et~al.}(1977)\citenamefont
  {Michele}, \citenamefont {P{\"a}tzold},\ and\ \citenamefont
  {Donis}}]{michele1977alignment}%
  \BibitemOpen
  \bibfield  {author} {\bibinfo {author} {\bibfnamefont {J.}~\bibnamefont
  {Michele}}, \bibinfo {author} {\bibfnamefont {R.}~\bibnamefont
  {P{\"a}tzold}},\ and\ \bibinfo {author} {\bibfnamefont {R.}~\bibnamefont
  {Donis}},\ }\bibfield  {title} {\bibinfo {title} {Alignment and aggregation
  effects in suspensions of spheres in non-{N}ewtonian media},\ }\href@noop {}
  {\bibfield  {journal} {\bibinfo  {journal} {Rheol. Acta}\ }\textbf {\bibinfo
  {volume} {16}},\ \bibinfo {pages} {317} (\bibinfo {year} {1977})}\BibitemShut
  {NoStop}%
\bibitem [{\citenamefont {D’Avino}\ \emph {et~al.}(2012)\citenamefont
  {D’Avino}, \citenamefont {Snijkers}, \citenamefont {Pasquino},
  \citenamefont {Hulsen}, \citenamefont {Greco}, \citenamefont {Maffettone},\
  and\ \citenamefont {Vermant}}]{d2012migration}%
  \BibitemOpen
  \bibfield  {author} {\bibinfo {author} {\bibfnamefont {G.}~\bibnamefont
  {D’Avino}}, \bibinfo {author} {\bibfnamefont {F.}~\bibnamefont {Snijkers}},
  \bibinfo {author} {\bibfnamefont {R.}~\bibnamefont {Pasquino}}, \bibinfo
  {author} {\bibfnamefont {M.~A.}\ \bibnamefont {Hulsen}}, \bibinfo {author}
  {\bibfnamefont {F.}~\bibnamefont {Greco}}, \bibinfo {author} {\bibfnamefont
  {P.~L.}\ \bibnamefont {Maffettone}},\ and\ \bibinfo {author} {\bibfnamefont
  {J.}~\bibnamefont {Vermant}},\ }\bibfield  {title} {\bibinfo {title}
  {Migration of a sphere suspended in viscoelastic liquids in {C}ouette flow:
  experiments and simulations},\ }\href@noop {} {\bibfield  {journal} {\bibinfo
   {journal} {Rheol. Acta}\ }\textbf {\bibinfo {volume} {51}},\ \bibinfo
  {pages} {215} (\bibinfo {year} {2012})}\BibitemShut {NoStop}%
\bibitem [{\citenamefont {Kim}\ \emph {et~al.}(2012)\citenamefont {Kim},
  \citenamefont {Ahn}, \citenamefont {Lee},\ and\ \citenamefont
  {Kim}}]{kim2012lateral}%
  \BibitemOpen
  \bibfield  {author} {\bibinfo {author} {\bibfnamefont {J.~Y.}\ \bibnamefont
  {Kim}}, \bibinfo {author} {\bibfnamefont {S.~W.}\ \bibnamefont {Ahn}},
  \bibinfo {author} {\bibfnamefont {S.~S.}\ \bibnamefont {Lee}},\ and\ \bibinfo
  {author} {\bibfnamefont {J.~M.}\ \bibnamefont {Kim}},\ }\bibfield  {title}
  {\bibinfo {title} {Lateral migration and focusing of colloidal particles and
  {DNA} molecules under viscoelastic flow},\ }\href@noop {} {\bibfield
  {journal} {\bibinfo  {journal} {Lab Chip}\ }\textbf {\bibinfo {volume}
  {12}},\ \bibinfo {pages} {2807} (\bibinfo {year} {2012})}\BibitemShut
  {NoStop}%
\bibitem [{\citenamefont {Van~Loon}\ \emph {et~al.}(2014)\citenamefont
  {Van~Loon}, \citenamefont {Fransaer}, \citenamefont {Clasen},\ and\
  \citenamefont {Vermant}}]{van2014string}%
  \BibitemOpen
  \bibfield  {author} {\bibinfo {author} {\bibfnamefont {S.}~\bibnamefont
  {Van~Loon}}, \bibinfo {author} {\bibfnamefont {J.}~\bibnamefont {Fransaer}},
  \bibinfo {author} {\bibfnamefont {C.}~\bibnamefont {Clasen}},\ and\ \bibinfo
  {author} {\bibfnamefont {J.}~\bibnamefont {Vermant}},\ }\bibfield  {title}
  {\bibinfo {title} {String formation in sheared suspensions in rheologically
  complex media: The essential role of shear thinning},\ }\href@noop {}
  {\bibfield  {journal} {\bibinfo  {journal} {J. Rheol.}\ }\textbf {\bibinfo
  {volume} {58}},\ \bibinfo {pages} {237} (\bibinfo {year} {2014})}\BibitemShut
  {NoStop}%
\bibitem [{\citenamefont {Lim}\ \emph {et~al.}(2014{\natexlab{a}})\citenamefont
  {Lim}, \citenamefont {Nam},\ and\ \citenamefont {Shin}}]{lim2014lateral}%
  \BibitemOpen
  \bibfield  {author} {\bibinfo {author} {\bibfnamefont {H.}~\bibnamefont
  {Lim}}, \bibinfo {author} {\bibfnamefont {J.}~\bibnamefont {Nam}},\ and\
  \bibinfo {author} {\bibfnamefont {S.}~\bibnamefont {Shin}},\ }\bibfield
  {title} {\bibinfo {title} {Lateral migration of particles suspended in
  viscoelastic fluids in a microchannel flow},\ }\href@noop {} {\bibfield
  {journal} {\bibinfo  {journal} {Microfluid. Nanofluid.}\ }\textbf {\bibinfo
  {volume} {17}},\ \bibinfo {pages} {683} (\bibinfo {year}
  {2014}{\natexlab{a}})}\BibitemShut {NoStop}%
\bibitem [{\citenamefont {Lim}\ \emph {et~al.}(2014{\natexlab{b}})\citenamefont
  {Lim}, \citenamefont {Ober}, \citenamefont {Edd}, \citenamefont {Desai},
  \citenamefont {Neal}, \citenamefont {Bong}, \citenamefont {Doyle},
  \citenamefont {McKinley},\ and\ \citenamefont {Toner}}]{lim2014inertio}%
  \BibitemOpen
  \bibfield  {author} {\bibinfo {author} {\bibfnamefont {E.~J.}\ \bibnamefont
  {Lim}}, \bibinfo {author} {\bibfnamefont {T.~J.}\ \bibnamefont {Ober}},
  \bibinfo {author} {\bibfnamefont {J.~F.}\ \bibnamefont {Edd}}, \bibinfo
  {author} {\bibfnamefont {S.~P.}\ \bibnamefont {Desai}}, \bibinfo {author}
  {\bibfnamefont {D.}~\bibnamefont {Neal}}, \bibinfo {author} {\bibfnamefont
  {K.~W.}\ \bibnamefont {Bong}}, \bibinfo {author} {\bibfnamefont {P.~S.}\
  \bibnamefont {Doyle}}, \bibinfo {author} {\bibfnamefont {G.~H.}\ \bibnamefont
  {McKinley}},\ and\ \bibinfo {author} {\bibfnamefont {M.}~\bibnamefont
  {Toner}},\ }\bibfield  {title} {\bibinfo {title} {Inertio-elastic focusing of
  bioparticles in microchannels at high throughput},\ }\href@noop {} {\bibfield
   {journal} {\bibinfo  {journal} {Nat. Commun.}\ }\textbf {\bibinfo {volume}
  {5}},\ \bibinfo {pages} {4120} (\bibinfo {year}
  {2014}{\natexlab{b}})}\BibitemShut {NoStop}%
\bibitem [{\citenamefont {Seo}\ \emph {et~al.}(2014)\citenamefont {Seo},
  \citenamefont {Byeon}, \citenamefont {Huh},\ and\ \citenamefont
  {Lee}}]{seo2014particle}%
  \BibitemOpen
  \bibfield  {author} {\bibinfo {author} {\bibfnamefont {K.~W.}\ \bibnamefont
  {Seo}}, \bibinfo {author} {\bibfnamefont {H.~J.}\ \bibnamefont {Byeon}},
  \bibinfo {author} {\bibfnamefont {H.~K.}\ \bibnamefont {Huh}},\ and\ \bibinfo
  {author} {\bibfnamefont {S.~J.}\ \bibnamefont {Lee}},\ }\bibfield  {title}
  {\bibinfo {title} {Particle migration and single-line particle focusing in
  microscale pipe flow of viscoelastic fluids},\ }\href@noop {} {\bibfield
  {journal} {\bibinfo  {journal} {RSC Adv.}\ }\textbf {\bibinfo {volume} {4}},\
  \bibinfo {pages} {3512} (\bibinfo {year} {2014})}\BibitemShut {NoStop}%
\bibitem [{\citenamefont {Li}\ \emph {et~al.}(2015)\citenamefont {Li},
  \citenamefont {McKinley},\ and\ \citenamefont {Ardekani}}]{li2015dynamics}%
  \BibitemOpen
  \bibfield  {author} {\bibinfo {author} {\bibfnamefont {G.}~\bibnamefont
  {Li}}, \bibinfo {author} {\bibfnamefont {G.~H.}\ \bibnamefont {McKinley}},\
  and\ \bibinfo {author} {\bibfnamefont {A.~M.}\ \bibnamefont {Ardekani}},\
  }\bibfield  {title} {\bibinfo {title} {Dynamics of particle migration in
  channel flow of viscoelastic fluids},\ }\href@noop {} {\bibfield  {journal}
  {\bibinfo  {journal} {J. Fluid Mech.}\ }\textbf {\bibinfo {volume} {785}},\
  \bibinfo {pages} {486} (\bibinfo {year} {2015})}\BibitemShut {NoStop}%
\bibitem [{\citenamefont {Lu}\ \emph {et~al.}(2017)\citenamefont {Lu},
  \citenamefont {Liu}, \citenamefont {Hu},\ and\ \citenamefont
  {Xuan}}]{lu2017particle}%
  \BibitemOpen
  \bibfield  {author} {\bibinfo {author} {\bibfnamefont {X.}~\bibnamefont
  {Lu}}, \bibinfo {author} {\bibfnamefont {C.}~\bibnamefont {Liu}}, \bibinfo
  {author} {\bibfnamefont {G.}~\bibnamefont {Hu}},\ and\ \bibinfo {author}
  {\bibfnamefont {X.}~\bibnamefont {Xuan}},\ }\bibfield  {title} {\bibinfo
  {title} {Particle manipulations in non-{N}ewtonian microfluidics: A review},\
  }\href@noop {} {\bibfield  {journal} {\bibinfo  {journal} {J. Colloid
  Interface Sci.}\ }\textbf {\bibinfo {volume} {500}},\ \bibinfo {pages} {182}
  (\bibinfo {year} {2017})}\BibitemShut {NoStop}%
\bibitem [{\citenamefont {Jaensson}\ \emph {et~al.}(2018)\citenamefont
  {Jaensson}, \citenamefont {Mitrias}, \citenamefont {Hulsen},\ and\
  \citenamefont {Anderson}}]{jaensson2018shear}%
  \BibitemOpen
  \bibfield  {author} {\bibinfo {author} {\bibfnamefont {N.~O.}\ \bibnamefont
  {Jaensson}}, \bibinfo {author} {\bibfnamefont {C.}~\bibnamefont {Mitrias}},
  \bibinfo {author} {\bibfnamefont {M.~A.}\ \bibnamefont {Hulsen}},\ and\
  \bibinfo {author} {\bibfnamefont {P.~D.}\ \bibnamefont {Anderson}},\
  }\bibfield  {title} {\bibinfo {title} {Shear-induced migration of rigid
  particles near an interface between a {N}ewtonian and a viscoelastic fluid},\
  }\href@noop {} {\bibfield  {journal} {\bibinfo  {journal} {Langmuir}\
  }\textbf {\bibinfo {volume} {34}},\ \bibinfo {pages} {1795} (\bibinfo {year}
  {2018})}\BibitemShut {NoStop}%
\bibitem [{\citenamefont {Yu}\ \emph {et~al.}(2019)\citenamefont {Yu},
  \citenamefont {Wang}, \citenamefont {Lin},\ and\ \citenamefont
  {Hu}}]{yu2019equilibrium}%
  \BibitemOpen
  \bibfield  {author} {\bibinfo {author} {\bibfnamefont {Z.}~\bibnamefont
  {Yu}}, \bibinfo {author} {\bibfnamefont {P.}~\bibnamefont {Wang}}, \bibinfo
  {author} {\bibfnamefont {J.}~\bibnamefont {Lin}},\ and\ \bibinfo {author}
  {\bibfnamefont {H.~H.}\ \bibnamefont {Hu}},\ }\bibfield  {title} {\bibinfo
  {title} {Equilibrium positions of the elasto-inertial particle migration in
  rectangular channel flow of {Oldroyd-B} viscoelastic fluids},\ }\href@noop {}
  {\bibfield  {journal} {\bibinfo  {journal} {J. Fluid Mech.}\ }\textbf
  {\bibinfo {volume} {868}},\ \bibinfo {pages} {316} (\bibinfo {year}
  {2019})}\BibitemShut {NoStop}%
\bibitem [{\citenamefont {Zhou}\ and\ \citenamefont
  {Papautsky}(2020)}]{zhou2020viscoelastic}%
  \BibitemOpen
  \bibfield  {author} {\bibinfo {author} {\bibfnamefont {J.}~\bibnamefont
  {Zhou}}\ and\ \bibinfo {author} {\bibfnamefont {I.}~\bibnamefont
  {Papautsky}},\ }\bibfield  {title} {\bibinfo {title} {Viscoelastic
  microfluidics: Progress and challenges},\ }\href@noop {} {\bibfield
  {journal} {\bibinfo  {journal} {Microsyst. Nanoeng.}\ }\textbf {\bibinfo
  {volume} {6}},\ \bibinfo {pages} {113} (\bibinfo {year} {2020})}\BibitemShut
  {NoStop}%
\bibitem [{\citenamefont {Del~Giudice}\ \emph {et~al.}(2017)\citenamefont
  {Del~Giudice}, \citenamefont {Haward},\ and\ \citenamefont
  {Shen}}]{del2017relaxation}%
  \BibitemOpen
  \bibfield  {author} {\bibinfo {author} {\bibfnamefont {F.}~\bibnamefont
  {Del~Giudice}}, \bibinfo {author} {\bibfnamefont {S.~J.}\ \bibnamefont
  {Haward}},\ and\ \bibinfo {author} {\bibfnamefont {A.~Q.}\ \bibnamefont
  {Shen}},\ }\bibfield  {title} {\bibinfo {title} {Relaxation time of dilute
  polymer solutions: A microfluidic approach},\ }\href@noop {} {\bibfield
  {journal} {\bibinfo  {journal} {J. Rheol.}\ }\textbf {\bibinfo {volume}
  {61}},\ \bibinfo {pages} {327} (\bibinfo {year} {2017})}\BibitemShut
  {NoStop}%
\bibitem [{\citenamefont {Vishnampet}\ and\ \citenamefont
  {Saintillan}(2012)}]{vishnampet2012concentration}%
  \BibitemOpen
  \bibfield  {author} {\bibinfo {author} {\bibfnamefont {R.}~\bibnamefont
  {Vishnampet}}\ and\ \bibinfo {author} {\bibfnamefont {D.}~\bibnamefont
  {Saintillan}},\ }\bibfield  {title} {\bibinfo {title} {Concentration
  instability of sedimenting spheres in a second-order fluid},\ }\href@noop {}
  {\bibfield  {journal} {\bibinfo  {journal} {Phys. Fluids}\ }\textbf {\bibinfo
  {volume} {24}} (\bibinfo {year} {2012})}\BibitemShut {NoStop}%
\bibitem [{\citenamefont {Einarsson}\ and\ \citenamefont
  {Mehlig}(2017)}]{einarsson2017spherical}%
  \BibitemOpen
  \bibfield  {author} {\bibinfo {author} {\bibfnamefont {J.}~\bibnamefont
  {Einarsson}}\ and\ \bibinfo {author} {\bibfnamefont {B.}~\bibnamefont
  {Mehlig}},\ }\bibfield  {title} {\bibinfo {title} {Spherical particle
  sedimenting in weakly viscoelastic shear flow},\ }\href@noop {} {\bibfield
  {journal} {\bibinfo  {journal} {Phys. Rev. Fluids}\ }\textbf {\bibinfo
  {volume} {2}},\ \bibinfo {pages} {063301} (\bibinfo {year}
  {2017})}\BibitemShut {NoStop}%
\bibitem [{\citenamefont {Zhang}\ \emph {et~al.}(2020)\citenamefont {Zhang},
  \citenamefont {Murch}, \citenamefont {Einarsson},\ and\ \citenamefont
  {Shaqfeh}}]{zhang2020lift}%
  \BibitemOpen
  \bibfield  {author} {\bibinfo {author} {\bibfnamefont {A.}~\bibnamefont
  {Zhang}}, \bibinfo {author} {\bibfnamefont {W.~L.}\ \bibnamefont {Murch}},
  \bibinfo {author} {\bibfnamefont {J.}~\bibnamefont {Einarsson}},\ and\
  \bibinfo {author} {\bibfnamefont {E.~S.~G.}\ \bibnamefont {Shaqfeh}},\
  }\bibfield  {title} {\bibinfo {title} {Lift and drag force on a spherical
  particle in a viscoelastic shear flow},\ }\href@noop {} {\bibfield  {journal}
  {\bibinfo  {journal} {J. Non-Newton. Fluid Mech.}\ }\textbf {\bibinfo
  {volume} {280}},\ \bibinfo {pages} {104279} (\bibinfo {year}
  {2020})}\BibitemShut {NoStop}%
\bibitem [{\citenamefont {Pak}\ \emph {et~al.}(2012)\citenamefont {Pak},
  \citenamefont {Zhu}, \citenamefont {Brandt},\ and\ \citenamefont
  {Lauga}}]{pak2012micropropulsion}%
  \BibitemOpen
  \bibfield  {author} {\bibinfo {author} {\bibfnamefont {O.~S.}\ \bibnamefont
  {Pak}}, \bibinfo {author} {\bibfnamefont {L.}~\bibnamefont {Zhu}}, \bibinfo
  {author} {\bibfnamefont {L.}~\bibnamefont {Brandt}},\ and\ \bibinfo {author}
  {\bibfnamefont {E.}~\bibnamefont {Lauga}},\ }\bibfield  {title} {\bibinfo
  {title} {Micropropulsion and microrheology in complex fluids via symmetry
  breaking},\ }\href@noop {} {\bibfield  {journal} {\bibinfo  {journal} {Phys.
  Fluids}\ }\textbf {\bibinfo {volume} {24}} (\bibinfo {year}
  {2012})}\BibitemShut {NoStop}%
\bibitem [{\citenamefont {Su}\ \emph {et~al.}(2022)\citenamefont {Su},
  \citenamefont {Castillo}, \citenamefont {Pak}, \citenamefont {Zhu},\ and\
  \citenamefont {Zenit}}]{su2022viscoelastic}%
  \BibitemOpen
  \bibfield  {author} {\bibinfo {author} {\bibfnamefont {Y.}~\bibnamefont
  {Su}}, \bibinfo {author} {\bibfnamefont {A.}~\bibnamefont {Castillo}},
  \bibinfo {author} {\bibfnamefont {O.~S.}\ \bibnamefont {Pak}}, \bibinfo
  {author} {\bibfnamefont {L.}~\bibnamefont {Zhu}},\ and\ \bibinfo {author}
  {\bibfnamefont {R.}~\bibnamefont {Zenit}},\ }\bibfield  {title} {\bibinfo
  {title} {Viscoelastic levitation},\ }\href@noop {} {\bibfield  {journal}
  {\bibinfo  {journal} {J. Fluid Mech.}\ }\textbf {\bibinfo {volume} {943}},\
  \bibinfo {pages} {A23} (\bibinfo {year} {2022})}\BibitemShut {NoStop}%
\bibitem [{\citenamefont {Kroo}\ \emph {et~al.}(2022)\citenamefont {Kroo},
  \citenamefont {Binagia}, \citenamefont {Eckman}, \citenamefont {Prakash},\
  and\ \citenamefont {Shaqfeh}}]{kroo2022freely}%
  \BibitemOpen
  \bibfield  {author} {\bibinfo {author} {\bibfnamefont {L.~A.}\ \bibnamefont
  {Kroo}}, \bibinfo {author} {\bibfnamefont {J.~P.}\ \bibnamefont {Binagia}},
  \bibinfo {author} {\bibfnamefont {N.}~\bibnamefont {Eckman}}, \bibinfo
  {author} {\bibfnamefont {M.}~\bibnamefont {Prakash}},\ and\ \bibinfo {author}
  {\bibfnamefont {E.~S.~G.}\ \bibnamefont {Shaqfeh}},\ }\bibfield  {title}
  {\bibinfo {title} {A freely suspended robotic swimmer propelled by
  viscoelastic normal stresses},\ }\href@noop {} {\bibfield  {journal}
  {\bibinfo  {journal} {J. Fluid Mech.}\ }\textbf {\bibinfo {volume} {944}},\
  \bibinfo {pages} {A20} (\bibinfo {year} {2022})}\BibitemShut {NoStop}%
\bibitem [{\citenamefont {Gagnon}\ \emph {et~al.}(2014)\citenamefont {Gagnon},
  \citenamefont {Keim}, \citenamefont {Shen},\ and\ \citenamefont
  {Arratia}}]{gagnon2014fluid}%
  \BibitemOpen
  \bibfield  {author} {\bibinfo {author} {\bibfnamefont {D.~A.}\ \bibnamefont
  {Gagnon}}, \bibinfo {author} {\bibfnamefont {N.~C.}\ \bibnamefont {Keim}},
  \bibinfo {author} {\bibfnamefont {X.}~\bibnamefont {Shen}},\ and\ \bibinfo
  {author} {\bibfnamefont {P.~E.}\ \bibnamefont {Arratia}},\ }\bibfield
  {title} {\bibinfo {title} {Fluid-induced propulsion of rigid particles in
  wormlike micellar solutions},\ }\href@noop {} {\bibfield  {journal} {\bibinfo
   {journal} {Phys. Fluids}\ }\textbf {\bibinfo {volume} {26}} (\bibinfo {year}
  {2014})}\BibitemShut {NoStop}%
\bibitem [{\citenamefont {Bird}\ \emph {et~al.}(1987)\citenamefont {Bird},
  \citenamefont {Armstrong},\ and\ \citenamefont
  {Hassager}}]{bird1987dynamics}%
  \BibitemOpen
  \bibfield  {author} {\bibinfo {author} {\bibfnamefont {R.~B.}\ \bibnamefont
  {Bird}}, \bibinfo {author} {\bibfnamefont {R.~C.}\ \bibnamefont
  {Armstrong}},\ and\ \bibinfo {author} {\bibfnamefont {O.}~\bibnamefont
  {Hassager}},\ }\href@noop {} {\emph {\bibinfo {title} {Dynamics of
  {P}olymeric {L}iquids. {Vol}. 1: Fluid {M}echanics}}}\ (\bibinfo  {publisher}
  {John Wiley and Sons Inc., New York, NY},\ \bibinfo {year}
  {1987})\BibitemShut {NoStop}%
\bibitem [{\citenamefont {Larson}(2013)}]{larson2013constitutive}%
  \BibitemOpen
  \bibfield  {author} {\bibinfo {author} {\bibfnamefont {R.~G.}\ \bibnamefont
  {Larson}},\ }\href@noop {} {\emph {\bibinfo {title} {Constitutive Equations
  for Polymer Melts and Solutions: {B}utterworths Series in Chemical
  Engineering}}}\ (\bibinfo  {publisher} {Butterworth-Heinemann},\ \bibinfo
  {year} {2013})\BibitemShut {NoStop}%
\bibitem [{\citenamefont {Morozov}\ and\ \citenamefont
  {Spagnolie}(2015)}]{Morozov2015}%
  \BibitemOpen
  \bibfield  {author} {\bibinfo {author} {\bibfnamefont {A.}~\bibnamefont
  {Morozov}}\ and\ \bibinfo {author} {\bibfnamefont {S.~E.}\ \bibnamefont
  {Spagnolie}},\ }\bibinfo {title} {Introduction to complex fluids},\ in\
  \href@noop {} {\emph {\bibinfo {booktitle} {Complex Fluids in Biological
  Systems: Experiment, Theory, and Computation}}},\ \bibinfo {editor} {edited
  by\ \bibinfo {editor} {\bibfnamefont {S.~E.}\ \bibnamefont {Spagnolie}}}\
  (\bibinfo  {publisher} {Springer New York},\ \bibinfo {address} {New York,
  NY},\ \bibinfo {year} {2015})\ pp.\ \bibinfo {pages} {3--52}\BibitemShut
  {NoStop}%
\bibitem [{\citenamefont {Apostolakis}\ \emph {et~al.}(2002)\citenamefont
  {Apostolakis}, \citenamefont {Mavrantzas},\ and\ \citenamefont
  {Beris}}]{apostolakis2002stress}%
  \BibitemOpen
  \bibfield  {author} {\bibinfo {author} {\bibfnamefont {M.~V.}\ \bibnamefont
  {Apostolakis}}, \bibinfo {author} {\bibfnamefont {V.~G.}\ \bibnamefont
  {Mavrantzas}},\ and\ \bibinfo {author} {\bibfnamefont {A.~N.}\ \bibnamefont
  {Beris}},\ }\bibfield  {title} {\bibinfo {title} {Stress gradient-induced
  migration effects in the {T}aylor--{C}ouette flow of a dilute polymer
  solution},\ }\href@noop {} {\bibfield  {journal} {\bibinfo  {journal} {J.
  Non-Newtonian Fluid Mech.}\ }\textbf {\bibinfo {volume} {102}},\ \bibinfo
  {pages} {409} (\bibinfo {year} {2002})}\BibitemShut {NoStop}%
\bibitem [{\citenamefont {Masoud}\ and\ \citenamefont
  {Stone}(2019)}]{masoud2019reciprocal}%
  \BibitemOpen
  \bibfield  {author} {\bibinfo {author} {\bibfnamefont {H.}~\bibnamefont
  {Masoud}}\ and\ \bibinfo {author} {\bibfnamefont {H.~A.}\ \bibnamefont
  {Stone}},\ }\bibfield  {title} {\bibinfo {title} {The reciprocal theorem in
  fluid dynamics and transport phenomena},\ }\href@noop {} {\bibfield
  {journal} {\bibinfo  {journal} {J. Fluid Mech.}\ }\textbf {\bibinfo {volume}
  {879}},\ \bibinfo {pages} {P1} (\bibinfo {year} {2019})}\BibitemShut
  {NoStop}%
\bibitem [{\citenamefont {Aubert}\ and\ \citenamefont
  {Tirrell}(1980)}]{aubert1980macromolecules}%
  \BibitemOpen
  \bibfield  {author} {\bibinfo {author} {\bibfnamefont {J.~H.}\ \bibnamefont
  {Aubert}}\ and\ \bibinfo {author} {\bibfnamefont {M.}~\bibnamefont
  {Tirrell}},\ }\bibfield  {title} {\bibinfo {title} {Macromolecules in
  nonhomogeneous velocity gradient fields},\ }\href@noop {} {\bibfield
  {journal} {\bibinfo  {journal} {J. Chem. Phys.}\ }\textbf {\bibinfo {volume}
  {72}},\ \bibinfo {pages} {2694} (\bibinfo {year} {1980})}\BibitemShut
  {NoStop}%
\bibitem [{\citenamefont {Helfand}\ and\ \citenamefont
  {Fredrickson}(1989)}]{helfand1989large}%
  \BibitemOpen
  \bibfield  {author} {\bibinfo {author} {\bibfnamefont {E.}~\bibnamefont
  {Helfand}}\ and\ \bibinfo {author} {\bibfnamefont {G.~H.}\ \bibnamefont
  {Fredrickson}},\ }\bibfield  {title} {\bibinfo {title} {Large fluctuations in
  polymer solutions under shear},\ }\href@noop {} {\bibfield  {journal}
  {\bibinfo  {journal} {Phys. Rev. Lett.}\ }\textbf {\bibinfo {volume} {62}},\
  \bibinfo {pages} {2468} (\bibinfo {year} {1989})}\BibitemShut {NoStop}%
\bibitem [{\citenamefont {Doi}(1990)}]{doi1990effects}%
  \BibitemOpen
  \bibfield  {author} {\bibinfo {author} {\bibfnamefont {M.}~\bibnamefont
  {Doi}},\ }\bibfield  {title} {\bibinfo {title} {Effects of viscoelasticity on
  polymer diffusion},\ }in\ \href@noop {} {\emph {\bibinfo {booktitle}
  {Dynamics and Patterns in Complex Fluids: New Aspects of the
  Physics-Chemistry Interface}}}\ (\bibinfo  {publisher} {Springer},\ \bibinfo
  {year} {1990})\ pp.\ \bibinfo {pages} {100--112}\BibitemShut {NoStop}%
\bibitem [{\citenamefont {Milner}(1991)}]{milner1991hydrodynamics}%
  \BibitemOpen
  \bibfield  {author} {\bibinfo {author} {\bibfnamefont {S.~T.}\ \bibnamefont
  {Milner}},\ }\bibfield  {title} {\bibinfo {title} {Hydrodynamics of
  semidilute polymer solutions},\ }\href@noop {} {\bibfield  {journal}
  {\bibinfo  {journal} {Phys. Rev. Lett.}\ }\textbf {\bibinfo {volume} {66}},\
  \bibinfo {pages} {1477} (\bibinfo {year} {1991})}\BibitemShut {NoStop}%
\bibitem [{\citenamefont {Mavrantzas}\ and\ \citenamefont
  {Beris}(1992)}]{mavrantzas1992modeling}%
  \BibitemOpen
  \bibfield  {author} {\bibinfo {author} {\bibfnamefont {V.~G.}\ \bibnamefont
  {Mavrantzas}}\ and\ \bibinfo {author} {\bibfnamefont {A.~N.}\ \bibnamefont
  {Beris}},\ }\bibfield  {title} {\bibinfo {title} {Modeling of the rheology
  and flow-induced concentration changes in polymer solutions},\ }\href@noop {}
  {\bibfield  {journal} {\bibinfo  {journal} {Phys. Rev. Lett.}\ }\textbf
  {\bibinfo {volume} {69}},\ \bibinfo {pages} {273} (\bibinfo {year}
  {1992})}\BibitemShut {NoStop}%
\bibitem [{\citenamefont {Tsouka}\ \emph {et~al.}(2014)\citenamefont {Tsouka},
  \citenamefont {Dimakopoulos}, \citenamefont {Mavrantzas},\ and\ \citenamefont
  {Tsamopoulos}}]{tsouka2014stress}%
  \BibitemOpen
  \bibfield  {author} {\bibinfo {author} {\bibfnamefont {S.}~\bibnamefont
  {Tsouka}}, \bibinfo {author} {\bibfnamefont {Y.}~\bibnamefont
  {Dimakopoulos}}, \bibinfo {author} {\bibfnamefont {V.}~\bibnamefont
  {Mavrantzas}},\ and\ \bibinfo {author} {\bibfnamefont {J.}~\bibnamefont
  {Tsamopoulos}},\ }\bibfield  {title} {\bibinfo {title} {Stress-gradient
  induced migration of polymers in corrugated channels},\ }\href@noop {}
  {\bibfield  {journal} {\bibinfo  {journal} {J. Rheol.}\ }\textbf {\bibinfo
  {volume} {58}},\ \bibinfo {pages} {911} (\bibinfo {year} {2014})}\BibitemShut
  {NoStop}%
\bibitem [{\citenamefont {Gomez-Solano}\ and\ \citenamefont
  {Bechinger}(2015)}]{gomez2015transient}%
  \BibitemOpen
  \bibfield  {author} {\bibinfo {author} {\bibfnamefont {J.~R.}\ \bibnamefont
  {Gomez-Solano}}\ and\ \bibinfo {author} {\bibfnamefont {C.}~\bibnamefont
  {Bechinger}},\ }\bibfield  {title} {\bibinfo {title} {Transient dynamics of a
  colloidal particle driven through a viscoelastic fluid},\ }\href@noop {}
  {\bibfield  {journal} {\bibinfo  {journal} {New J. Phys.}\ }\textbf {\bibinfo
  {volume} {17}},\ \bibinfo {pages} {103032} (\bibinfo {year}
  {2015})}\BibitemShut {NoStop}%
\bibitem [{\citenamefont {Teran}\ \emph {et~al.}(2010)\citenamefont {Teran},
  \citenamefont {Fauci},\ and\ \citenamefont
  {Shelley}}]{teran2010viscoelastic}%
  \BibitemOpen
  \bibfield  {author} {\bibinfo {author} {\bibfnamefont {J.}~\bibnamefont
  {Teran}}, \bibinfo {author} {\bibfnamefont {L.}~\bibnamefont {Fauci}},\ and\
  \bibinfo {author} {\bibfnamefont {M.}~\bibnamefont {Shelley}},\ }\bibfield
  {title} {\bibinfo {title} {Viscoelastic fluid response can increase the speed
  and efficiency of a free swimmer},\ }\href@noop {} {\bibfield  {journal}
  {\bibinfo  {journal} {Phys. Rev. Lett.}\ }\textbf {\bibinfo {volume} {104}},\
  \bibinfo {pages} {038101} (\bibinfo {year} {2010})}\BibitemShut {NoStop}%
\bibitem [{\citenamefont {Liu}\ \emph {et~al.}(2011)\citenamefont {Liu},
  \citenamefont {Powers},\ and\ \citenamefont {Breuer}}]{liu2011force}%
  \BibitemOpen
  \bibfield  {author} {\bibinfo {author} {\bibfnamefont {B.}~\bibnamefont
  {Liu}}, \bibinfo {author} {\bibfnamefont {T.~R.}\ \bibnamefont {Powers}},\
  and\ \bibinfo {author} {\bibfnamefont {K.~S.}\ \bibnamefont {Breuer}},\
  }\bibfield  {title} {\bibinfo {title} {Force-free swimming of a model helical
  flagellum in viscoelastic fluids},\ }\href@noop {} {\bibfield  {journal}
  {\bibinfo  {journal} {Proc. Natl. Acad. Sci. USA}\ }\textbf {\bibinfo
  {volume} {108}},\ \bibinfo {pages} {19516} (\bibinfo {year}
  {2011})}\BibitemShut {NoStop}%
\bibitem [{\citenamefont {Oppenheimer}\ \emph {et~al.}(2016)\citenamefont
  {Oppenheimer}, \citenamefont {Navardi},\ and\ \citenamefont
  {Stone}}]{oppenheimer2016motion}%
  \BibitemOpen
  \bibfield  {author} {\bibinfo {author} {\bibfnamefont {N.}~\bibnamefont
  {Oppenheimer}}, \bibinfo {author} {\bibfnamefont {S.}~\bibnamefont
  {Navardi}},\ and\ \bibinfo {author} {\bibfnamefont {H.~A.}\ \bibnamefont
  {Stone}},\ }\bibfield  {title} {\bibinfo {title} {Motion of a hot particle in
  viscous fluids},\ }\href@noop {} {\bibfield  {journal} {\bibinfo  {journal}
  {Phys. Rev. Fluids}\ }\textbf {\bibinfo {volume} {1}},\ \bibinfo {pages}
  {014001} (\bibinfo {year} {2016})}\BibitemShut {NoStop}%
\bibitem [{\citenamefont {Liebchen}\ \emph {et~al.}(2018)\citenamefont
  {Liebchen}, \citenamefont {Monderkamp}, \citenamefont {Ten~Hagen},\ and\
  \citenamefont {L{\"o}wen}}]{liebchen2018viscotaxis}%
  \BibitemOpen
  \bibfield  {author} {\bibinfo {author} {\bibfnamefont {B.}~\bibnamefont
  {Liebchen}}, \bibinfo {author} {\bibfnamefont {P.}~\bibnamefont
  {Monderkamp}}, \bibinfo {author} {\bibfnamefont {B.}~\bibnamefont
  {Ten~Hagen}},\ and\ \bibinfo {author} {\bibfnamefont {H.}~\bibnamefont
  {L{\"o}wen}},\ }\bibfield  {title} {\bibinfo {title} {Viscotaxis:
  Microswimmer navigation in viscosity gradients},\ }\href@noop {} {\bibfield
  {journal} {\bibinfo  {journal} {Phys. Rev. Lett.}\ }\textbf {\bibinfo
  {volume} {120}},\ \bibinfo {pages} {208002} (\bibinfo {year}
  {2018})}\BibitemShut {NoStop}%
\bibitem [{\citenamefont {Datt}\ and\ \citenamefont
  {Elfring}(2019)}]{datt2019active}%
  \BibitemOpen
  \bibfield  {author} {\bibinfo {author} {\bibfnamefont {C.}~\bibnamefont
  {Datt}}\ and\ \bibinfo {author} {\bibfnamefont {G.~J.}\ \bibnamefont
  {Elfring}},\ }\bibfield  {title} {\bibinfo {title} {Active particles in
  viscosity gradients},\ }\href@noop {} {\bibfield  {journal} {\bibinfo
  {journal} {Phys. Rev. Lett.}\ }\textbf {\bibinfo {volume} {123}},\ \bibinfo
  {pages} {158006} (\bibinfo {year} {2019})}\BibitemShut {NoStop}%
\bibitem [{\citenamefont {Dandekar}\ and\ \citenamefont
  {Ardekani}(2020)}]{dandekar2020swimming}%
  \BibitemOpen
  \bibfield  {author} {\bibinfo {author} {\bibfnamefont {R.}~\bibnamefont
  {Dandekar}}\ and\ \bibinfo {author} {\bibfnamefont {A.~M.}\ \bibnamefont
  {Ardekani}},\ }\bibfield  {title} {\bibinfo {title} {Swimming sheet in a
  viscosity-stratified fluid},\ }\href@noop {} {\bibfield  {journal} {\bibinfo
  {journal} {J. Fluid Mech.}\ }\textbf {\bibinfo {volume} {895}},\ \bibinfo
  {pages} {R2} (\bibinfo {year} {2020})}\BibitemShut {NoStop}%
\bibitem [{\citenamefont {Shaik}\ and\ \citenamefont
  {Elfring}(2021)}]{shaik2021hydrodynamics}%
  \BibitemOpen
  \bibfield  {author} {\bibinfo {author} {\bibfnamefont {V.~A.}\ \bibnamefont
  {Shaik}}\ and\ \bibinfo {author} {\bibfnamefont {G.~J.}\ \bibnamefont
  {Elfring}},\ }\bibfield  {title} {\bibinfo {title} {Hydrodynamics of active
  particles in viscosity gradients},\ }\href@noop {} {\bibfield  {journal}
  {\bibinfo  {journal} {Phys. Rev. Fluids}\ }\textbf {\bibinfo {volume} {6}},\
  \bibinfo {pages} {103103} (\bibinfo {year} {2021})}\BibitemShut {NoStop}%
\bibitem [{\citenamefont {Esparza~L{\'o}pez}\ \emph {et~al.}(2021)\citenamefont
  {Esparza~L{\'o}pez}, \citenamefont {Gonzalez-Gutierrez}, \citenamefont
  {Solorio-Ordaz}, \citenamefont {Lauga},\ and\ \citenamefont
  {Zenit}}]{esparza2021dynamics}%
  \BibitemOpen
  \bibfield  {author} {\bibinfo {author} {\bibfnamefont {C.}~\bibnamefont
  {Esparza~L{\'o}pez}}, \bibinfo {author} {\bibfnamefont {J.}~\bibnamefont
  {Gonzalez-Gutierrez}}, \bibinfo {author} {\bibfnamefont {F.}~\bibnamefont
  {Solorio-Ordaz}}, \bibinfo {author} {\bibfnamefont {E.}~\bibnamefont
  {Lauga}},\ and\ \bibinfo {author} {\bibfnamefont {R.}~\bibnamefont {Zenit}},\
  }\bibfield  {title} {\bibinfo {title} {Dynamics of a helical swimmer crossing
  viscosity gradients},\ }\href@noop {} {\bibfield  {journal} {\bibinfo
  {journal} {Phys. Rev. Fluids}\ }\textbf {\bibinfo {volume} {6}},\ \bibinfo
  {pages} {083102} (\bibinfo {year} {2021})}\BibitemShut {NoStop}%
\bibitem [{\citenamefont {Stehnach}\ \emph {et~al.}(2021)\citenamefont
  {Stehnach}, \citenamefont {Waisbord}, \citenamefont {Walkama},\ and\
  \citenamefont {Guasto}}]{stehnach2021viscophobic}%
  \BibitemOpen
  \bibfield  {author} {\bibinfo {author} {\bibfnamefont {M.~R.}\ \bibnamefont
  {Stehnach}}, \bibinfo {author} {\bibfnamefont {N.}~\bibnamefont {Waisbord}},
  \bibinfo {author} {\bibfnamefont {D.~M.}\ \bibnamefont {Walkama}},\ and\
  \bibinfo {author} {\bibfnamefont {J.~S.}\ \bibnamefont {Guasto}},\ }\bibfield
   {title} {\bibinfo {title} {Viscophobic turning dictates microalgae transport
  in viscosity gradients},\ }\href@noop {} {\bibfield  {journal} {\bibinfo
  {journal} {Nat. Phys.}\ }\textbf {\bibinfo {volume} {17}},\ \bibinfo {pages}
  {926} (\bibinfo {year} {2021})}\BibitemShut {NoStop}%
\bibitem [{\citenamefont {Olsen}\ \emph {et~al.}(2021)\citenamefont {Olsen},
  \citenamefont {Angheluta},\ and\ \citenamefont
  {Flekk{\o}y}}]{olsen2021active}%
  \BibitemOpen
  \bibfield  {author} {\bibinfo {author} {\bibfnamefont {K.~S.}\ \bibnamefont
  {Olsen}}, \bibinfo {author} {\bibfnamefont {L.}~\bibnamefont {Angheluta}},\
  and\ \bibinfo {author} {\bibfnamefont {E.~G.}\ \bibnamefont {Flekk{\o}y}},\
  }\bibfield  {title} {\bibinfo {title} {Active {Brownian} particles moving
  through disordered landscapes},\ }\href@noop {} {\bibfield  {journal}
  {\bibinfo  {journal} {Soft Matter}\ }\textbf {\bibinfo {volume} {17}},\
  \bibinfo {pages} {2151} (\bibinfo {year} {2021})}\BibitemShut {NoStop}%
\bibitem [{\citenamefont {De~Corato}\ and\ \citenamefont
  {Mart{\'\i}nez-Lera}(2025)}]{de2025enhanced}%
  \BibitemOpen
  \bibfield  {author} {\bibinfo {author} {\bibfnamefont {M.}~\bibnamefont
  {De~Corato}}\ and\ \bibinfo {author} {\bibfnamefont {P.}~\bibnamefont
  {Mart{\'\i}nez-Lera}},\ }\bibfield  {title} {\bibinfo {title} {Enhanced
  rotational diffusion and spontaneous rotation of an active {J}anus disk in a
  complex fluid},\ }\href@noop {} {\bibfield  {journal} {\bibinfo  {journal}
  {Soft Matter}\ }\textbf {\bibinfo {volume} {21}},\ \bibinfo {pages} {186}
  (\bibinfo {year} {2025})}\BibitemShut {NoStop}%
\end{thebibliography}

%

\clearpage
\section{End Matter}
Here, we report the steady-state dimensionless equations and elaborate the asymptotic analysis. 
\subsection{Oldroyd-B (OB) and Giesekus models}
Using the characteristic scales defined in the main text, we obtain the dimensionless governing equations. The momentum balance and continuity equations are
\begin{equation} \label{dimless_mombalcont}
    \grad^* \cdot \left( 2 \beta \boldsymbol{D}^* -p^*\boldsymbol{I} + \boldsymbol{\tau}^*  \right)= \boldsymbol{0} \, \, \, ; \, \, \,   \grad^* \cdot \boldsymbol{v}^* =0\,\,.
\end{equation}
The OB model for the polymeric stress is
\begin{equation}
    De  \, \overset{\nabla}{\boldsymbol{\tau}^*} + \boldsymbol{\tau}^* =2(1-\beta) \,\boldsymbol{D}^* \, \, , 
\end{equation}
and the Giesekus counterpart is
\begin{equation}
    De  \, \overset{\nabla}{\boldsymbol{\tau}^*} + \boldsymbol{\tau}^*+ \frac{\alpha \, De}{1-\beta} \, \boldsymbol{\tau}^*\cdot \boldsymbol{\tau}^* =2(1-\beta) \,\boldsymbol{D}^* \, \, , 
\end{equation}
where $\alpha$ must be between 0 and 1 \cite{larson2013constitutive}. The force-free condition is given by
\begin{equation} \label{dimless_forcefree}
    \int_{\mS}  
    \left( 2 \beta \boldsymbol{D}^* -p^*\boldsymbol{I} + \boldsymbol{\tau}^*  \right): \boldsymbol{n} \boldsymbol{e}_x d \mS = 0 \, \,.
\end{equation}
The velocity far from the particle is 
\begin{equation}\label{dimless_bcfarfield}
    \boldsymbol{v}^* = - \boldsymbol{e}_z - V_x^* \boldsymbol{e}_x\,\, ,
\end{equation}
and that at the particle surface reads
\begin{equation}\label{dimless_bcpart}
    \boldsymbol{v}^* = \omeganon \boldsymbol{e}_y \times \boldsymbol{r}^* \, \, .
\end{equation}

\textit{Perturbation expansion}---We seek a solution
in the form of a perturbation expansion for small $De$ values, expanding the relevant variables as $\boldsymbol{v}^*=\boldsymbol{v}_{0}^*+ De \,  \boldsymbol{v}_{1}^* + \mathcal{O}(De^2)$, $p^*=p_{0}^*+ De \,  p_{1}^* + \mathcal{O}(De^2)$, $\boldsymbol{\tau}^*=\boldsymbol{\tau}_{0}^*+ De \, \boldsymbol{\tau}_{1}^* + \mathcal{O}(De^2)$, and $V_x^*= De \,  V_{x,1}^* + \mathcal{O}(De^2)$.

\textit{Zeroth-order problem}---The zero-order equations are obtained by substituting the perturbation expansion into the main equations and retaining only the terms that do not depend on $De$. 
The polymeric stress is $\boldsymbol{\tau}_{0}^* = 2(1-\beta) \,\boldsymbol{D}^*_0$. The zeroth-order momentum and continuity equations are 
\begin{equation}\label{zero_mombal}
    \grad^* \cdot \left( 2  \boldsymbol{D}^*_0 -p^*_0\boldsymbol{I}  \right)= \boldsymbol{0} \, \, \, \, ; \, \,\, \,    
    \grad^* \cdot \boldsymbol{v}^*_0 =0\,\,,
\end{equation}
corresponding to the Newtonian Stokes flow.
Its flow field can be obtained by superimposing the flow past a sphere and that around a spinning sphere, resulting in
\begin{multline}\label{zero_sol}
\boldsymbol{v}_0^* = \boldsymbol{e}_z \cdot \left[ \left(\frac{3}{4 r^*} +\frac{1}{4{r^*}^3}\right)\boldsymbol{I} +\frac{3}{4}\left( \frac{1}{{r^*}^3}-\frac{1}{{r^*}^5}\right)\boldsymbol{r}^*\boldsymbol{r}^* \right]+\\
\omeganon \boldsymbol{e}_y \times \boldsymbol{r}^*/{r^*}^3 - \boldsymbol{e}_z\,\,. 
\end{multline}

\textit{First-order problem}---The first-order equations are obtained by substituting the perturbation expansion into the equations and retaining only the terms proportional to $De$.
The first order solution of the polymeric stress is 
\begin{equation}
    \boldsymbol{\tau}_1  =2(1-\beta) \,\boldsymbol{D}_1^*- \overset{\nabla}{\boldsymbol{\tau}_0^*} \, \, .
\end{equation}
for the OB fluid
and 
\begin{equation}
    \boldsymbol{\tau}_1  =2(1-\beta) \,\boldsymbol{D}_1^*- \overset{\nabla}{\boldsymbol{\tau}_0^*}-\frac{\alpha \, De}{1-\beta} \boldsymbol{\tau}_0^* \cdot \boldsymbol{\tau}_0^*  \, \, ,
\end{equation}
for the Giesekus model.
The momentum and the continuity equations are
\begin{equation}
    \grad \cdot \left( 2 \beta \boldsymbol{D}_1^* -p_1^*\boldsymbol{I} + \boldsymbol{\tau}_1^*  \right)= \boldsymbol{0} \, \, \, \, ; \,\,\,\,
    \grad \cdot \boldsymbol{v}_1^* =0 \, \, .
\end{equation}
The far-field velocity $   \boldsymbol{v}_1^* = - V_{x,1}^* \boldsymbol{e}_x $ and  $\boldsymbol{v}_1^* =  \boldsymbol{0}$ at the particle surface.
The first-order force-free condition is $\int_{\mS}  \left( 2 \beta \boldsymbol{D}^*_1 -p^*_1 \boldsymbol{I} + \boldsymbol{\tau}^*_1  \right): \boldsymbol{n}\boldsymbol{e}_x \, d\mS= 0$.

\textit{Generalized reciprocal theorem}---Using the Lorentz  reciprocal theorem~\cite{leal1979motion,masoud2019reciprocal}, we compute the first-order velocity $V_{x,1}^*$ by the volume integral over the fluid domain, $\mV$.
For the OB model, 
\begin{equation} \label{OB_GRT}
    V_{x,1}^* = \frac{1}{6\pi} \int_{\mV} \grad\hat{\boldsymbol{v}}^* :  \overset{\nabla}{\boldsymbol{\tau}_0^*} \, d\mV \,\, ,
\end{equation}
where $\hat{\boldsymbol{v}}^{*}$ is an auxiliary velocity field corresponding to Newtonian Stokes flow with an ambient velocity $1\boldsymbol{e}_x$ past a stationary sphere. Likewise, for the Giesekus model,
\begin{equation}\label{GIE_GRT}
    V_{x,1}^* = \frac{1}{6\pi} \int_{\mV} \grad\hat{\boldsymbol{v}}^* :  \left(\overset{\nabla}{\boldsymbol{\tau}_0^*} -\frac{\alpha \, De}{1-\beta} \boldsymbol{\tau}_0^* \cdot \boldsymbol{\tau}_0^* \right)\, d\mV \,\,.
\end{equation}

\subsection{Stress-gradient induced polymer transport}
The OB and Giesekus models assume a uniform far-field number density $n=n_{\infty}$. Here, we extend the OB model by incorporating a nonuniform density $n$. Taking $n_\infty$ as the characteristic scale, the dimensionless number density $n^*=n/n_{\infty}$.
While the momentum and continuity equations remain unchanged, the polymeric stress follows an adapted constitutive law~\cite{apostolakis2002stress,tsouka2014stress}
\begin{equation}
    De  \, \overset{\nabla}{\boldsymbol{\tau}^*} + \boldsymbol{\tau}^*  + (1-\beta) \left(\frac{D n^*}{ D t^* }\right)\, \boldsymbol{I} =2(1-\beta) \, n^* \,\boldsymbol{D}^*\,\,,
\end{equation}
where $D ()/Dt^*$ denotes the material derivative. Specifically, the spatio-temporal evolution of $n^*$ is governed  by 
\begin{equation} \label{adv_diff_pol}
    Pe \,  \frac{D \, n^*}{D \, t} =\grad \cdot \left( \grad n^* - \frac{De}{1-\beta}\grad \cdot \boldsymbol{\tau}^* \right)\,\,.
\end{equation}
The force-free condition and boundary conditions for the velocity are the same as Eqs. \labelcref{dimless_forcefree,dimless_bcfarfield,dimless_bcpart}. The boundary condition for $n^*$ at the particle surface reads
\begin{equation}
    \boldsymbol{n} \cdot \left( -\grad n^* + \frac{De}{1-\beta}\grad \cdot \boldsymbol{\tau}^* \right) =0\,\,,
\end{equation}
and that far from the particle is $n^*=1$.

\textit{Perturbation expansion}---We seek a solution in the form of a perturbation expansion for small $De$ and $Pe$. 
We find that all asymptotic terms proportional to powers of $Pe$ only are zero. 
The variables are expanded as: $\boldsymbol{v}^*=\boldsymbol{v}_{0}^*+ De \,  \boldsymbol{v}_{1}^* + \mathcal{O}(De^2, De Pe)$, $p^*=p_{0}^*+ De \,  p_{1}^* + \mathcal{O}(De^2, De Pe)$, $\boldsymbol{\tau}^*=\boldsymbol{\tau}_{0}^*+ De \, \boldsymbol{\tau}_{1}^* + \mathcal{O}(De^2, De Pe)$, $n^*=1+ De \, n_{1}^* + \mathcal{O}(De^2, De Pe)$, and $V_x^*= De \,  V_{x,1}^* + \mathcal{O}(De^2, De Pe)$. 

\textit{Zeroth-order problem}---The zeroth-order equations and their solutions, $\boldsymbol{v}^*_0$ and $p_0$, are the same as those for the OB and Giesekus models 
because the zeroth-order polymer density is uniformly equal to its far-field value. 

\textit{First-order problem}---The first order  polymeric stress takes the form, 
\begin{equation}\label{eq:tau1}
    \boldsymbol{\tau}_1^*  =2(1-\beta) \,\left( n_1^* \,\boldsymbol{D}_0^*+\,\boldsymbol{D}_1^*\right)- \overset{\nabla}{\boldsymbol{\tau}_0^*} -(1-\beta) \left(\frac{D n_1^*}{ D t }\right)\, \boldsymbol{I}\, \, .
\end{equation}
We note that the last term is an isotropic tensor 
not contributing to the velocity field. It can be absorbed by defining a modified $P_1^* = p_1^* +(1-\beta)D n_1^*/ D t  $, yielding
\begin{equation}
    \boldsymbol{\tau}_1^*  =2(1-\beta) \,\left( n_1^* \,\boldsymbol{D}_0^*+\,\boldsymbol{D}_1^*\right)- \overset{\nabla}{\boldsymbol{\tau}_0^*} \, \, .
\end{equation}
The momentum and continuity equations are $\grad \cdot \left( 2 \beta \boldsymbol{D}_1^* -P_1^*\boldsymbol{I} + \boldsymbol{\tau}_1^*  \right)= \boldsymbol{0} \, \, \, \, \, ; \, \, \, \, \, \grad \cdot \boldsymbol{v}_1^* =0$.
The velocity far from the particle is $\boldsymbol{v}_1^* = - V_{x,1}^* \boldsymbol{e}_x$ and at the particle surface $\boldsymbol{v}_1^* =  \boldsymbol{0}$.
The force-free condition is $\int_{S^*} \left( 2 \beta \boldsymbol{D}_1^* -P_1^*\boldsymbol{I} + \boldsymbol{\tau}_1^*  \right): \boldsymbol{n} \boldsymbol{xe}_x dS^*= 0$.
To proceed, we first solve for the number density $n_1^*$. Its steady-state solution follows: 
\begin{equation}
     \grad \cdot \left( \grad n_1^* - \frac{1}{1-\beta}\grad \cdot \boldsymbol{\tau}_0^* \right) =0 \, \, ,
\end{equation}
which can reduced to the Laplace equation      $\nabla^2 n_1^* =0$ 
realizing $\grad \boldsymbol{\nabla}: \boldsymbol{\tau}_0^* =0$.
The zero-flux boundary condition at the particle surface becomes 
\begin{equation}
    \boldsymbol{n} \cdot \left( -\grad n_1^* + \frac{1}{1-\beta}\grad \cdot \boldsymbol{\tau}_0^* \right) =0 \, \, ,
\end{equation}
and $n_1^*=0$ in the far field. We obtain $n_1^*=3 \, z^*/2 {r^*}^3$ that is independent of $\omeganon$, implying that 
the particle rotation does not cause the polymer inhomogeneity at the first order.

\textit{Generalized reciprocal theorem}---Employing again the reciprocal theorem, we derive the first-order velocity as
\begin{equation}
    V_{x,1}^* = \frac{1}{6\pi} \int_{\mV} \grad\hat{\boldsymbol{v}}^* : \left[ 2(1-\beta) \, n_1^* \,\boldsymbol{D}_0^*+ \overset{\nabla}{\boldsymbol{\tau}_0^*} \right] \, d\mV\,\,.
\end{equation}
Noting that the integral involving $\overset{\nabla}{\boldsymbol{\tau}_0^*} $ is identical to that for the OB model [see Eq. \eqref{OB_GRT}] and is thus zero, the first-order velocity is given by
\begin{equation}
    V_{x,1}^* = \frac{1}{6\pi} \int_{\mV} \grad\hat{\boldsymbol{v}}^* :\left[ 2(1-\beta) \, n_1^* \,\boldsymbol{D}_0^*\right] \, d\mV \, \, ,
\end{equation}
which crucially depends on the first-order inhomogeneous polymer distribution $n_1^*$ derived above. 

\subsection{Numerical methods}
We solve the dimensionless equations using the FEM package, COMSOL Multiphysics (I-Math, Singapore), specifically employing its Partial Differential Equation Interface. Exploiting the mirror-symmetry about the $x^*z^*$-plane, we consider a fluid domain in the form of a semi-spherical shell with an inner radius of $1$ and an outer radius of $200$. The inner and outer boundaries represent the particle surface and far field, respectively. 
The domain is discretized by $\approx 198,000$ tetrahedral elements, with polynomial order two for velocity $\boldsymbol{u}^*$, and order one for the other variables (pressure $p^*$, polymeric stress $\boldsymbol{\tau}^*$, and number density $n^*$), respectively. The mesh is refined near the particle surface, yielding a minimum element size of $0.1$. We perform a mesh-independence study, as summarized in Table~\ref{tab:my-table}.

To stabilize the simulations, streamline diffusion is applied to the $\boldsymbol{\tau}^*$ and $n^*$ equations. 
Imposing the force-free condition to compute the migration velocity of particles is achieved through a `Global Equations' node.
We solve the equations in a fully coupled manner, utilizing the direct solver MUMPS.

\begin{table}[hpt!]
\centering
\begin{tabular}{ l | l | l}
\thead{Minimum \\ element size}  & \thead{Total number \\ of elements}   & \thead{Migration \\ velocity $-V_x^*$ }   \\
\hline
 0.2   &   $\approx 136,000$     &    0.0252    \\
\hline
0.135       &   $\approx 165,000$     &   0.0287     \\
\hline
0.1   \cellcolor[gray]{0.8}   & $\approx 198,000$  \cellcolor[gray]{0.8}      &    0.0296   \cellcolor[gray]{0.8}  \\
\hline
0.07       &   $\approx 262,000$      &  0.0298 \\
\hline
0.05 & $\approx 359,000$ &  0.0297 \\ 
\end{tabular}
\caption{Mesh dependence evaluated at $\beta=0.1$, $De=Pe=0.3$, and $\omeganon=0.5$. The mesh used in our study is highlighted in gray.}
\label{tab:my-table}
\end{table}

\subsection{Rheological properties of the micellar solution}
For the theory-experiment comparison shown in Figure~\ref{fig:comparison}, 
the rheological fluid tested by \citet{cao2023} was a wormlike micellar solution composed of about 5 mM equimolar cetylpyridinium chloride monohydrate (CPyCl) and sodium salicylate (NaSal) dissolved in water. 
Its rheological properties were measured by the same group~\cite{gomez2015transient}. As reported, its zero-shear viscosity is $0.045\pm 0.005\, \text{Pa}\cdot \text{s} $, substantially larger than the viscosity of water, thereby leading to $\beta \rightarrow 0$. The inset of Figure 6(c) in Ref.~\cite{gomez2015transient}  indicates that the solution's largest relaxation time is approximately $0.9\,\text{s}$, which we choose as the relaxation time $\lambda$ in our constitutive model.

\end{document}